\pdfoutput=1
\documentclass[journal]{IEEEtran}
\usepackage[T1]{fontenc}
\usepackage{tikz}
\usepackage{epstopdf}
\usepackage{graphicx}
\usepackage{float}
\usepackage{url}
\usepackage{amsmath}
\usepackage[cmintegrals]{newtxmath}
\usepackage{booktabs}
\usepackage{pdflscape}
\usepackage{multirow}
\usepackage{subcaption}
\usepackage[normalem]{ulem}
\usepackage{caption}

\begin{document}

\title{Counter-Unmanned Aircraft System(s) (C-UAS): State of the Art, Challenges and Future Trends}

\author{Jian~Wang,
        Yongxin~Liu,
        and~Houbing~Song,~\IEEEmembership{Senior~Member,~IEEE} 
\thanks{Jian Wang, Yongxin Liu, and Houbing Song are with the Security and Optimization for Networked Globe Laboratory (SONG Lab, www.SONGLab.us), Department of Electrical Engineering and Computer Science, Embry-Riddle Aeronautical University, Daytona Beach, FL 32114 USA e-mail: WANGJ14@my.erau.edu; LIUY11@my.erau.edu; Houbing.Song@erau.edu}
\thanks{Manuscript received November 15, 2019; revised March 19, 2020.}
}

\markboth{IEEE AESS Systems Magazine} {Shell \MakeLowercase{\textit{et al.}}: Bare Demo of IEEEtran.cls for IEEE Communications Society Journals}
\maketitle
\begin{abstract}
Unmanned aircraft systems (UAS), or unmanned aerial vehicles (UAVs), often referred to as drones, have been experiencing healthy growth in the United States and around the world. The positive uses of UAS have the potential to save lives, increase safety and efficiency, and enable more effective science and engineering research. However, UAS are subject to threats stemming from increasing reliance on computer and communication technologies, which place public safety, national security, and individual privacy at risk. To promote safe, secure and privacy-respecting UAS operations, there is an urgent need for innovative technologies for detecting, tracking, identifying and mitigating UAS. A Counter-UAS (C-UAS) system is defined as a system or device capable of lawfully and safely disabling, disrupting, or seizing control of an unmanned aircraft or unmanned aircraft system. Over the past 5 years, significant research efforts have been made to detect, and mitigate UAS: detection technologies are based on acoustic, vision, passive radio frequency, radar, and data fusion; and mitigation technologies include physical capture or jamming. In this paper, we provide a comprehensive survey of existing literature in the area of C-UAS, identify the challenges in countering unauthorized or unsafe UAS, and evaluate the trends of detection and mitigation for protecting against UAS-based threats. The objective of this survey paper is to present a systematic introduction of C-UAS technologies, thus fostering a research community committed to the safe integration of UAS into the airspace system.

\end{abstract}
\begin{IEEEkeywords}
Unmanned Aircraft Systems, Unmanned Aerial Vehicles, Drones, Counter-Unmanned Aircraft System(s) (C-UAS), Detection, Sensing,  Mitigation, Safety, Security, Privacy, Software Defined Radios.
\end{IEEEkeywords}

\IEEEpeerreviewmaketitle
\section{Introduction}
\IEEEPARstart{A}n unmanned aircraft system is an unmanned aircraft (an aircraft that is operated without the possibility of direct human intervention from within or on the aircraft) and associated elements
(including communication links and the components that control the unmanned aircraft) that are required for the operator to operate safely and efficiently in the airspace system. Over the last 5 years, unmanned aircraft systems (UAS), or unmanned aerial vehicles (UAVs), often referred to as drones, have been experiencing healthy growth in the United States and around the world \cite{FAA}. According to the Federal Aviation Administration (FAA) aerospace forecast fiscal years 2019-2039, the model UAS fleet is set to grow from the present 1.25 million units to around 1.39 million units by 2023 and the non-model UAS fleet is set to grow from the present 277,000 aircraft to over 835,000 aircraft by 2023 \cite{FAA_UAS}. The positive uses of UAS have the potential to save lives, increase safety and efficiency, and enable more effective science and engineering research \cite{SC17}. These uses may include modelers experimenting with small UAS, performing numerous functions including aerial photography and personal recreational flying, commercial operators experimenting with package and medical supply delivery and providing support for search and rescue missions.

While the introduction of UAS in the airspace system has opened up numerous possibilities, UAS can also be used for malicious schemes by terrorists, criminal organizations (including transnational organizations), and lone actors with specific objectives. UAS-based threats stem from increasing reliance on computer and communication technologies, placing public safety, national security, and individual privacy at risk \cite{SP17}. 
\begin{itemize}
    \item \textit{Safety}: Unsafe UAS operations involve operating UAS near other aircraft, especially near airports; over groups of people, public events, or stadiums full of people; near emergencies such as fires or hurricane recovery efforts; or under the influence of drugs or alcohol \cite{UASpatent}\cite{COMMAG18}. Reports of UAS sightings from pilots, citizens and law enforcement have increased dramatically over the past five years \cite{Sightings}. A recent notable UAS incident was Gatwick Airport UAS incident: Between December 19 and 21, 2018, hundreds of flights were cancelled at Gatwick Airport near London, England, following reports of UAS sightings close to the runway. The reports caused major disruption, affecting approximately 140,000 passengers and 1,000 flights \cite{Gatwick}.
    \item \textit{Security}: Unsecure UAS operations involve operating UAS over designated national security sensitive facilities, such as military bases, national landmarks (such as the Statue of Liberty, Hoover Dam, Mt. Rushmore), and certain critical infrastructure (such as nuclear power plants), among others \cite{SecuritySensitive}\cite{DHS}. 
    \item \textit{Privacy}: Privacy-invading UAS operations involve operating UAS with their camera on when pointing inside a private residence \cite{SP17}\cite{carr2013unmanned}\cite{MobiSec20}.
\end{itemize}
To promote safe, secure and privacy-respecting UAS operations, there is an urgent need for innovative technologies for detecting, tracking, identifying and mitigating UAS. A counter-UAS (C-UAS) system is defined as a system or device capable of lawfully and safely disabling, disrupting, or seizing control of an unmanned aircraft or unmanned aircraft system \cite{Negation}. Typically such a system is comprised of two subsystems: one for detection and the other for mitigation \cite{DARPA-RFI}\cite{UASpatent}. The ideal UAS detection subsystem, will detect, track, identify an unmanned aircraft or unmanned aircraft system, have a small footprint, and support highly automated operations. The ideal UAS mitigation subsystem \cite{Negation} will lawfully and safely disable, disrupt, or seize control of an unmanned aircraft or unmanned aircraft system \cite{Negation}, while ensuring low collateral damage and low cost per engagement. 

Due to the fact that UAS are aircraft without a human pilot onboard that are controlled by an operator remotely or programmed to fly autonomously, protecting against UAS-based threats is very challenging. The challenges, which UAS can present to critical infrastructure and several courses of action that law enforcement and critical infrastructure owners and operators may want to take, include: (1) the pilots of the UAS are not able to receive commands from airspace authorities; (2) the pilots of the UAS commonly take actions according to video steams and GPS trajectories; (3) the communication links between pilots and UAS are vulnerable to interference.

Over the past 5 years, a lot of research efforts have been made to detect, track, identify and mitigate UAS: detection technologies are based on acoustic \cite{Ac6}, vision \cite{christnacher2016optical}, passive radio frequency (RF) \cite{RF8337905}, radar \cite{RADAR8384626} and data fusion \cite{Da3}; and mitigation technologies include physical capture (containment netting) \cite{Tomase2019Scalable}, jamming (RF command and control (C2) jamming and spoofing, or Global Positioning System (GPS) jamming and spoofing) \cite{8398711}, and destruction (RF C2 intercept and control) \cite{8100595}. However, these efforts are not mature: lack of scalability, modularity, or affordability. Innovative technologies towards relatively mature scalable, modular, and affordable approaches to detection and negation of UAS are desired. 

In this paper, we provide a comprehensive survey of existing literature in the area of UAS detection and negation, identify the challenges in countering adversary UAS, and evaluate the trends of detection and negation for protecting against UAS-based threats. The objective of this survey paper is to present a systematic introduction of C-UAS technologies, thus fostering a research community committed to the safe integration of UAS into the airspace system.

The remainder of this paper is structured as follows. Section II presents various UAS-based threats and introduces restrictions that commonly affect UAS flights. Sections III and IV presents the state of the art detection and mitigation, respectively. Section V identifies the challenges in countering unauthorized or unsafe UAS. Section VI evaluates the trends of detection and mitigation for protecting against UAS-based threats. Section VII concludes this paper.

\section{Background}
In this section, we discuss why we need UAS detection and mitigation, different types of UAS-based threats, and the common airspace restrictions applicable to UAS flights.

\subsection{Reported UAS Sightings}
\begin{figure}[htbp]
    \centering
    \includegraphics[width = 0.95\linewidth]{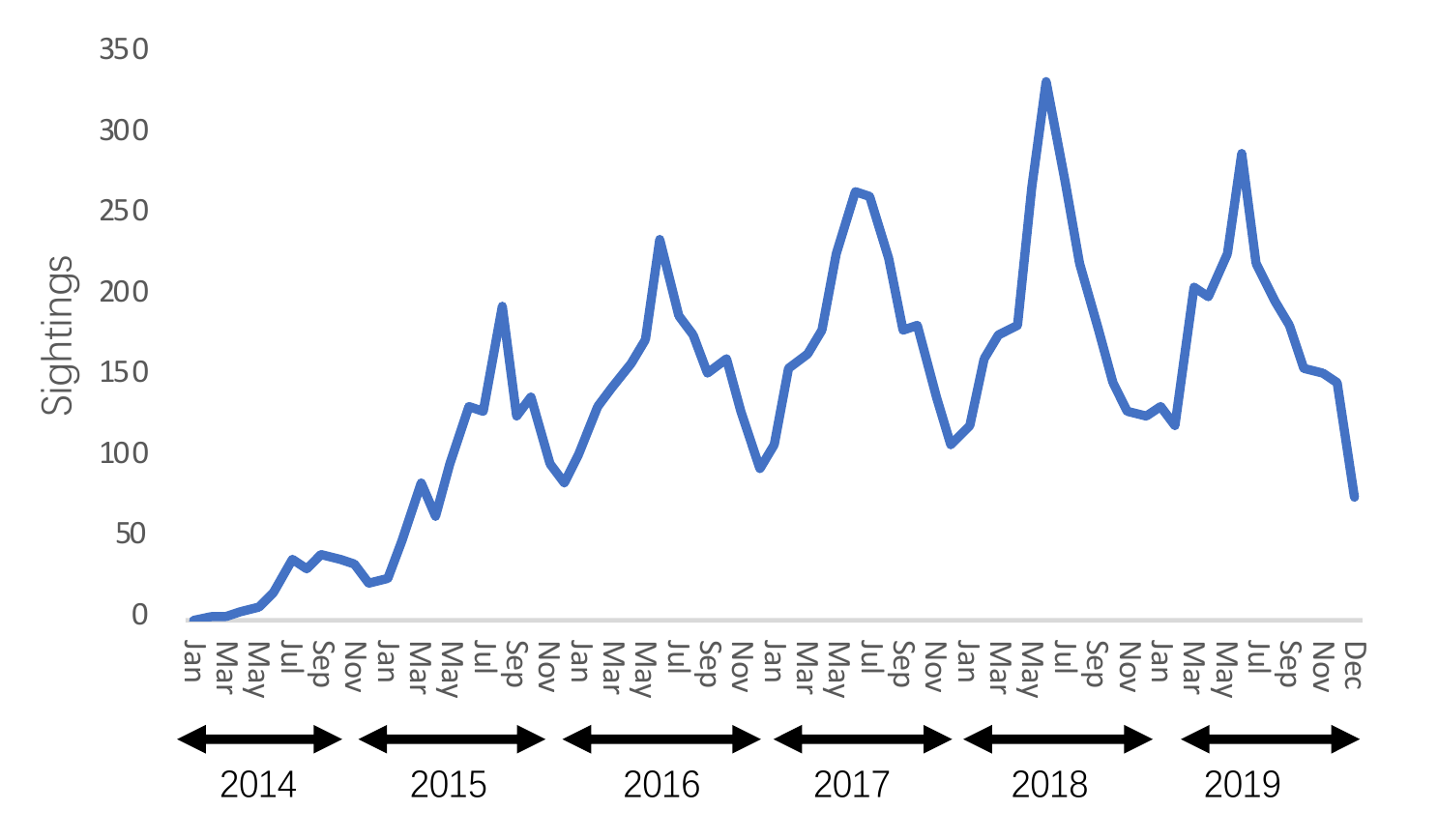}
    \caption{Temporal distribution of UAS sightings}
    \label{figTemporalDistributions}
\end{figure}

\begin{figure}[htbp]
    \centering
    \includegraphics[width =0.9\linewidth]{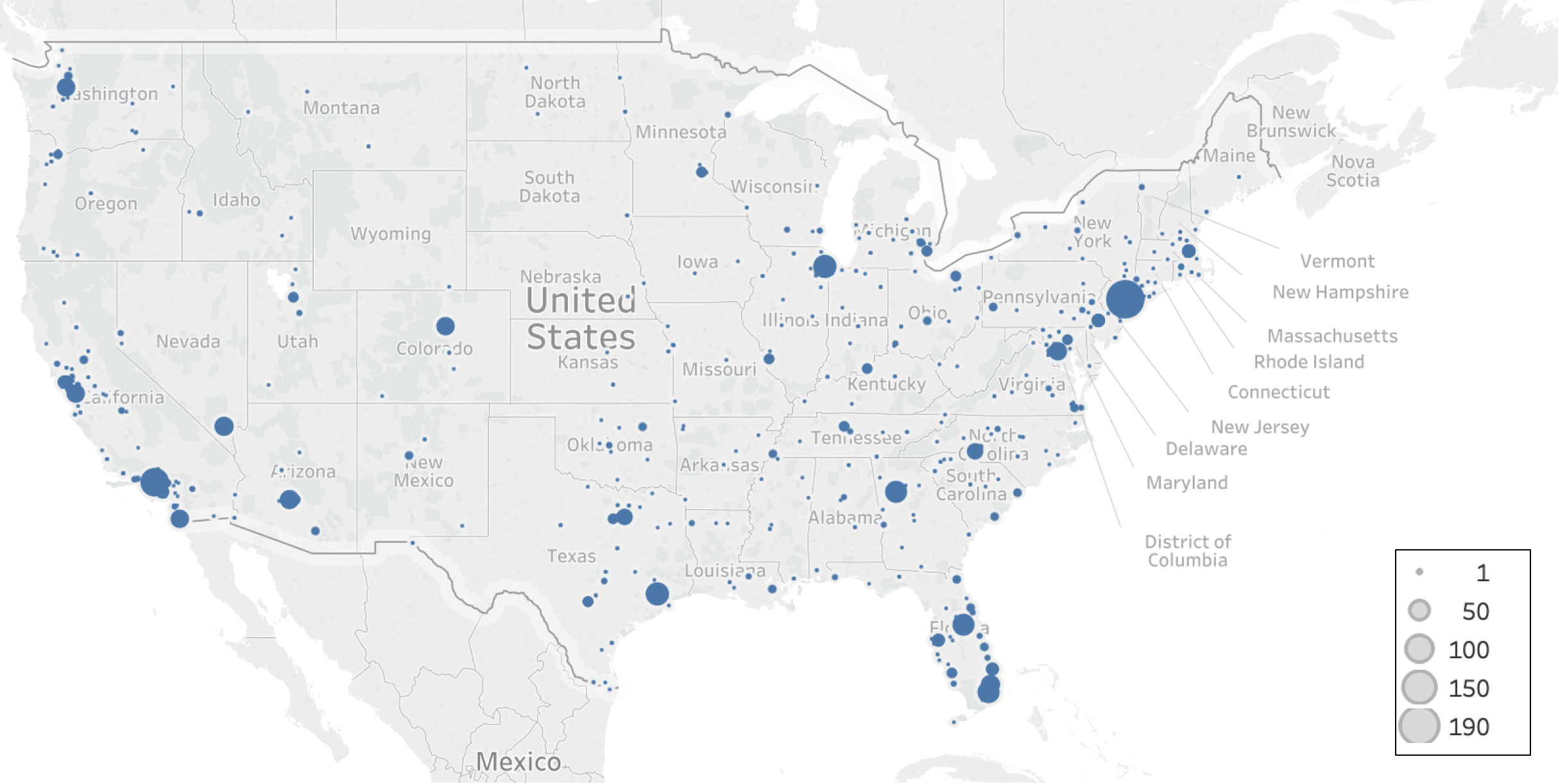}
    \caption{Spatial Distribution of UAS sightings}
    \label{figSpatialDistribution}
\end{figure}

Over the past five years, reports of UAS sightings from pilots, citizens and law enforcement have increased dramatically. Each month the FAA receives more than 100 such reports. The monthly UAS sightings between January 2014 and December 2019 are shown in Fig. \ref{figTemporalDistributions}, from which we could observe that unsafe and unauthorized UAS operations have been increasing dramatically. It is interesting that the most UAS sightings occur in the summer months. The UAS sightings in each state are shown in Fig. \ref{figSpatialDistribution}, from which we could observe that unsafe and unauthorized UAS operations occur in most populated US states. 

\subsection{UAS-based Threats}
In this paper, we classify the UAS-based threats into three categories: public safety, national security, and individual privacy, as shown in Table \ref{tab_SafetyThreats}. UAS-based public safety threats are due to operating UAS near other aircraft, especially near airports; over groups of people, public events, or stadiums full of people \cite{COMMAG18}; near emergencies such as fires or hurricane recovery efforts; or under the influence of drugs or alcohol; UAS-based national security threats are due to operating UAS over designated national security sensitive facilities, such as military bases, national landmarks, and certain critical infrastructure, among others \cite{SecuritySensitive, DHS}; UAS-based privacy threats are due to operating UAS with their camera on when pointing inside a private residence \cite{SP17,carr2013unmanned}. 
\begin{table*}
    \centering
    \caption{Classification of UAS-based Threats}
    \label{tab_SafetyThreats}
    \begin{tabular}{@{}llll@{}}
    \toprule
    Threats & Threatened entities & Threat mode & Consequences \\ \midrule
    Safety & Human, facilities and high value targets & \begin{tabular}[c]{@{}l@{}}Collisions, indirect hazards \\ and controlled attacks\end{tabular} & Injuries or damage of properties. \\
    Security & High value targets & \begin{tabular}[c]{@{}l@{}}Aerial imaging and \\ posterior reconstruction\end{tabular} & \begin{tabular}[c]{@{}l@{}}Disclosure of sensitive information \\ and national security issues\end{tabular} \\
    Privacy & Human & \begin{tabular}[c]{@{}l@{}}Aerial imaging or \\ real-time video stream\end{tabular} & Privacy invasion \\ \bottomrule
    \end{tabular}
\end{table*}

\begin{table*}
\centering
\caption{Some Recent UAS-based Threats}
\label{tab_TypicalEvents}
$
\begin{tabular}{p{2.2cm}p{4.8cm}p{1.1cm}p{1.1cm}p{2.5cm}p{3.5cm}}
\toprule

Event & Description & Time & Category & Location & Influence \\ 
\hline

Drone intrusion in non-flight zone & There were 54 drone incursions in Super Bowl \cite{giaritelli_2019} & April 10, 2019 & Safety & Mercedes-Benz Stadium, USA & Causing special attention of law enforcement agencies\\
Drones threatening aviation safety & Two drones crashed into landing zone when a B787-9 was landing. \cite{mowat_2019}& April 21, 2019 & Safety & Heathrow Airport, UK & RAF have spent \pounds 5 million to prevent future attacks \\

Drones stealing information & Spy drones hacked wireless networks with software defined radios. \cite{geek_com_2011} & Jul 29, 2011 & Security & Las Vegas, USA & Loss of information security. \\

Drones threatening national security & Border patrol spotted drones trying to help migrants illegally enter America \cite{norman_2019} & April 19, 2019 & Security & US-Mexico border, USA & Increasing border protection difficulty \\
Drones threatening personal privacy & Illegal activities obtained individual's privacy using drones. \cite{seabrook_valdes_2018} & Oct 28, 2018 & Privacy  & Orlando, USA & Police taking part in investigation \\
Drones threatening personal privacy & Residents disturbed by peeping drone outside bedroom window \cite{margaritoff_2018} & Feb 23, 2018 & Privacy & Upper Hutt, New Zealand & Privacy invasion concerns \\
\bottomrule
\end{tabular}$
\end{table*}
\begin{table*}
    \renewcommand{\arraystretch}{1.3}
    \caption{Airspace Restrictions Applicable to UAS by FAA (March 2020)}
    \label{tab_FAARestrict}
    \centering
    \newcommand{\tabincell}[2]{\begin{tabular}{@{}#1@{}}#2\end{tabular}}
    \begin{tabular}{p{3cm}p{10cm}} 
    \toprule
    Stadiums                    &Operations are prohibited within a radius of three nautical miles. Operations are prohibited starting one hour before and ending one hour after scheduled time of important events.\\
    Airports                    &1. One must have a Remote Pilot Certificate and get permission from Air traffic control (ATC). 2. Model aeroclub organization must notify the airport operator and control tower to fly within 5 miles. 3. Public entity (law enforcement or government agency) may apply for special permission. \\
    Security sensitive airspace &Operations are prohibited from the ground up to 400 feet above ground level.\\
    Capital areas              & Varying according to the policies of government, in Washington, DC., airspace is governed by a Special Flight Rules Area (SFRA), UAS operations are restricted within a 30-mile radius of Ronald Reagan Washington National Airport (15-mile inner ring: prohibitted without specific permission from FAA; 15 to 30 miles outer ring: registered, light and small UAV can fly lower than 400 feet and visual range in clear weather).\\
    Restricted or Special Use Airspace & Certain areas where drones and other aircraft are not permitted to fly without special permission, or where limitations must be imposed for any number of reasons. \\
    Temporary Flight Restriction (TFR)&A TFR defines a restricted airspace due to a hazardous condition or specific events. List of TFR can be found at \url{https://tfr.faa.gov/tfr2/list.html}\\
    Emergency and Rescue Operations&FAA prohibits drones over any emergency or rescue operations. It’s a federal crime to interfere with firefighting aircraft regardless of whether restrictions are established.\\
    \bottomrule
    \end{tabular}

\end{table*}

\subsubsection{Safety Threats} The unmanned nature of UAS operations raises two unique safety concerns that are not present in manned-aircraft operations: the pilot of the small UAS, who is physically separated from it during flight, may not have the ability to see manned aircraft in the air in time to prevent a mid-air collision, and the pilot of the small UAS could lose control of it due to a failure of the communications link between the small UAS and the pilot's handset for controlling the UAS \cite{GAO}. Safety risks related to the use of UAS include the potential for unintentional collisions between a small UAS and a manned aircraft or other objects, causing damage to property, or injury or death to persons \cite{GAO}. Operating UAS around airplanes, helicopters and airports is dangerous and illegal. 
\subsubsection{Security Threats}
UAS-related national security threats may include \cite{DHS-CI}:
\begin{itemize}
    \item Weaponized or Smuggling Payloads: Depending on power and payload size, UAS may be capable of transporting contraband, chemical, or other explosive/weaponized payloads.
\item Prohibited Surveillance and Reconnaissance: UAS are capable of silently monitoring a large area from the sky for nefarious purposes.
\item Intellectual Property Theft: UAS can be used to perform cyber crimes involving theft of trade secrets, technologies, or sensitive information.
\end{itemize}
\subsubsection{Privacy Threats} UAS-based privacy threats lie in intentional disruption or harassment. UAS may be used to disrupt or invade the privacy of other individuals \cite{DHS-CI}.

Some recent UAS-based threats are given in Table \ref{tab_TypicalEvents}, which shows the time and location of threat occurrence, threat category, and corresponding consequences. 

\subsection{Airspace Restrictions Applicable to UAS Flights}
In the United States, there are several types of airspace restrictions that commonly affect UAS flights \cite{AirspaceRestrictions}, as shown in Table \ref{tab_FAARestrict}. From the table, the most stringent restrictions are security sensitive airspace restrictions which prohibit UAS operations from the ground up to 400 feet above ground level, and apply to all types and purposes of UAS flight operations; for stadiums and sporting events, restrictions are valid during gathering of the crowds; Washington DC has the largest spatial restriction circle (30 miles). 

The FAA established a series of rules and regulations which apply to UAS operations, based on the type of UAS flier: recreational fliers and modeler community-based organizations, certificated remote pilots including commercial operators, public safety and government, and educational users \cite{FAA}. UAS that weigh more than 0.55 pounds must be registered with the FAA. In addition, flying UAS that are less than 55 pounds for work or business requires remote pilot certificates \cite{FAA}. 

\begin{figure*}
    \centering
    \includegraphics[scale = 0.65]{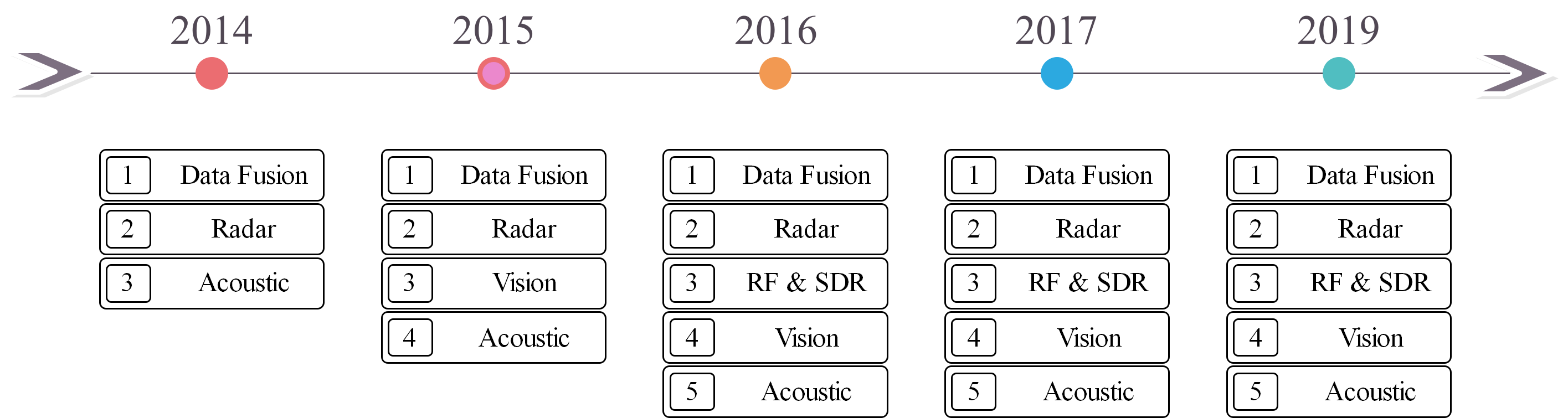}
    \caption{Evolution of UAS Detection Technologies}
    \label{fig1}
\end{figure*}

Although the FAA has established guidelines and regulations,  and reports of UAS sightings from pilots, citizens and law enforcement have increased dramatically over the past five years, as shown in Figures \ref{figTemporalDistributions} and \ref{figSpatialDistribution}. Therefore, there is an urgent need for promoting safe, secure and privacy-respecting UAS operations. We  envision  that  an  integrated system capable of detecting and negating UAS will be essential to the safe integration of UAS into the airspace system. Such a system needs novel technologies in the following two main areas:
\begin{itemize}
    \item Detection: The  ideal  UAS  detection  will  detect, track and identify an unmanned aircraft or unmanned aircraft system,  have  a small footprint, support highly automated operations and location functions.
    \item Mitigation: The ideal UAS mitigation system will lawfully and safely disable, disrupt, or seize control of an unmanned aircraft or unmanned aircraft system, while ensuring low collateral damage and low cost per engagement.
\end{itemize}
A survey of UAS detection and mitigation is the first step in leveraging communications and signal processing to develop low footprint UAS detection solutions, in terms of size, weight, power, and manning, as well as varied and low collateral damage UAS mitigation techniques and effectors, towards safe integration of UAS into the airspace system.

\section{State of the Art UAS Detection}
Since 2014, five technologies have been proposed for UAS detection, including acoustic, vision, passive radio frequency, radar, and data fusion. We summarized the evolution of UAS detection technologies in Fig.~\ref{fig1}, where different technologies are ranked based on the popularity for each year. From the figure, we know that data fusion approaches have been the most popular technology for UAS detection while acoustic based approaches have been the least popular. In this section, we will introduce each UAS detection technology and discuss its advantages and disadvantages.

\subsection{Acoustic based UAS Detection}
Acoustic based UAS detection leverages acoustic sensors to capture the sound of UAS, identify and track the UAS with audio. Acoustic sensor arrays, which are deployed around the restricted areas, record the audio signal periodically and deliver the audio signal to the ground stations. The ground stations extract the features of the audio signal to determine whether the UAS are approaching.

Conventionally, after receiving the audio signal of UAS, the power spectrum or frequency spectrum will be analyzed to identify the UAS. Vil\'imek, J. et al. adopted the linear predictive coding to distinguish the sound of UAS engine from the sound of car engine but the performance is subject to weather conditions \cite{Ac2}. Kim, J. et al. designed a real-time UAS sound detection and analysis system which could acquire the real-time sound data from the sensor and recognize the UAS \cite{Ac6}. Jang, B. et al applied the euclidean distance and Scale-Invariant Feature Transform (SIFT) to distinguish the UAV engine sound from the background sound and demonstrated their effectiveness, even though the power spectrum of the noise is larger than that of the UAS sound. However, in practice, their processing efficiency is poor \cite{AC3}.

Due to light weight, low-cost and easy assembly, acoustic sensors could be used to construct acoustic acquiring array and deployed in the target area to locate and track the trajectories of the UAS. An approach is proposed to deploy acoustic sensor array which acquires the sound of engine in UAS. The acoustic sensor array consists of 24 custom-built microphones which locate and track the UAS collaboratively. They calibrated each sensor with Time Delay Of Arrival (TDOA) and predicted the UAS flight path with beamforming. They could track the flight path well but this approach could not work well in a large scale space, and the accuracy of the sensor is dependent seriously on calibration \cite{Ac1}. Two arrays consists of 4 microphone sensors to improve the capability of the UAS localization. Due to the multipath effect, they provided a Gauss prior probability density function to improve the TDOA estimation. Their arrays could be deployed in specific area efficiently and achieved good performance to track the UAS. However, their system is not stable when it works for a long time \cite{Ac4}. Advanced acoustic cameras are leveraged to detect and track the UAS. To be specific, they used 2 to 4 acoustic cameras to capture the strength distribution of sound and fused the strength distribution to compute the location of the UAS indoors and outdoors \cite{Ac8}. An audio-assisted camera array is deployed to detect the UAS, which captured the video and audio signals at the same time and classified the object with Histogram of Oriented Gradient (HOG) feature and Mel Frequency Cepstral Coefficents (MFCC) feature \cite{Ac7}. 

Different from the above conventional methods, significant research efforts have been made to leverage machine learning to classify the UAS from audio data.Support Vector Machine (SVM) is implemented to analyze the mid term signal of UAS engine and constructed the signal fingerprint of UAS. Their results showed that the classifier could precisely distinguish the UAS in some scenarios \cite{Ac5}. An approach is proposed to transform the detection of the presence of UAS to a binary classification problem, and used Gaussian Mixture Model (GMM), Convolutional Neural Network (CNN) and Recurrent Neural Network (RNN) to detect UAS. Their results showed that could work well with the short input signal within 240ms \cite{Ac9}. 

The current acoustic based UAS detection technologies can recognize and locate the UAS precisely to meet the accuracy requirement of UAS detection. However, the nature of acoustic approaches limits the deployment and detection of UAS in a large scale. Machine learning (ML) presents a tremendous opportunity of integrating with the acoustic sensing into acoustic based UAS detection for improved UAS detection performance. 

\subsection{Passive RF based Detection}
\begin{figure}[H]
    \centering
    \includegraphics[width =\linewidth]{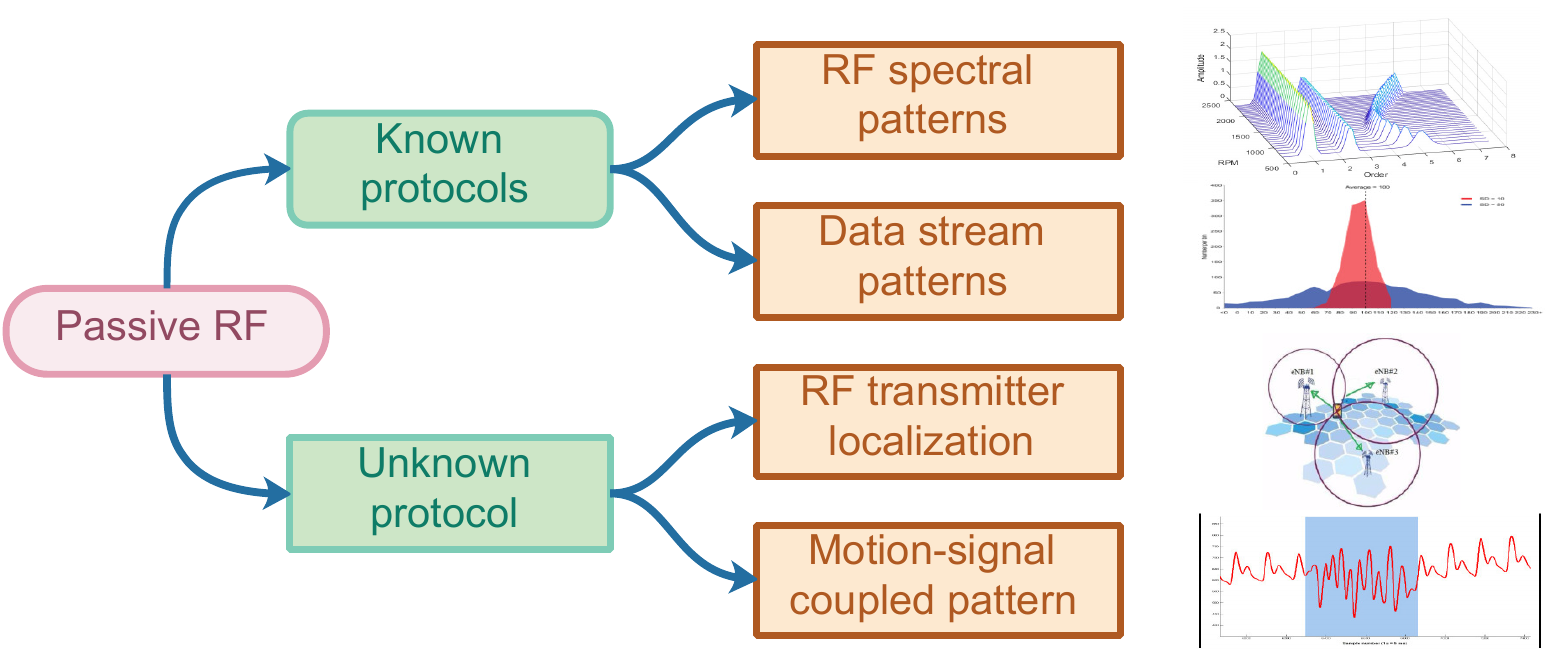}
    \caption{Categorization of passive RF approaches}
    \label{figPassiveRFcategory}
\end{figure}

UAS usually maintain at least one Radio Frequency (RF) communication data link to its remote controller to either receive control commands or deliver aerial images. In this case, the spectral patterns of such transmission are used as an important evidence for the detection and localization of UAS. Different passive RF technologies are shown in Fig. \ref{figPassiveRFcategory}. In most cases, Software Defined Radio (SDR) receivers are employed to intercept the RF channels. 

To utilize the spectrum patterns of UAS, in \cite{RF8353152}, an Artificial Neural Networks (ANN) detection algorithm for a UAS RF signal was proposed which employs three signal features: improved slope, improved skewness, and improved kurtosis. It was shown that the proposed algorithm based on ANN outperforms other recognition technologies of the improved slope, skewness, and kurtosis of signal spectrum. 
Data traffic patterns are also an important feature to specify UAS. In \cite{RF8457352}, a UAS detection and identification system, which utilized commercial off-the-shelf hardware to passively listen to the wireless signal between UAS and their controllers for packet transmission characteristics, was proposed. They mainly extracted the packet length distribution of UAS and evaluated the prototype system with three types of UAS. Their experiment results demonstrated the feasibility of using the data frame length to identify different UAS within 20 seconds. The increasing amount of commercial UAS, using WiFi as control and First Person View (FPV) video streaming protocol, motivated the method in \cite{RF8337905}, a UAS detection approach based on WiFi fingerprint. The method identified the presence of unauthorized UAS in a nearby area by monitoring the data traffic. 

Data traffic based methods or pure spectrum pattern methods depends highly on the telemetry protocol or RF front-ends of UAS. These methods may not be able to identify UAS operating with unknown telemetry protocols. Therefore, some researches focus on the use of the coupled kinetics motion patterns of flying UAS from their radio signal to specify their presence. In \cite{RF8205982} and \cite{RFNguyen2017Matthan}, Matthan, a cost-effective and passive RF-based UAS detection system was introduced. Their system detected the presence of UAS by identifying the unique signatures of its coupled vibration and shifting patterns in the transmitted wireless signals. The joint detector integrated evidence from both a frequency-based detector that indicated UAS body vibration as well as a wavelet-based detector that captured the sudden shifts of drone frame by computing wavelets at different scales from the temporal RF signal. A similar approach was also presented in \cite{RFnguyen2016investigating}. 

Localization is also an important part of passive RF based UAS detection. In \cite{RADAR8398721}, the authors divided the complete detection procedure for UAS detection into RF spectrum sensing and the Direction of Arrival (DoA) estimated. \cite{RF8484827} presented a useful experiment, which used commercial off the shelf Field Programmable Gate Array (FPGA) based SDR system for detecting and locating small UAS. Their results demonstrated that it is possible to develop a UAS detection system capable of detecting small UAS with error of 50-75m using commercial FPGA-based SDR system. The SDR system with optimized clock synchronization would radically reduce the measurement error in distance. How to implement robust localization algorithms with acceptable accuracy on ubiquitous hardware is still an open problem in passive RF detection of UAS. How to deploy such passive RF based systems on the ground stations or other UAS platforms is another open problem in the detection of UAS.

\subsection{Vision based UAS Detection}
\begin{figure}[H]
    \centering
    \includegraphics[width =0.95\linewidth]{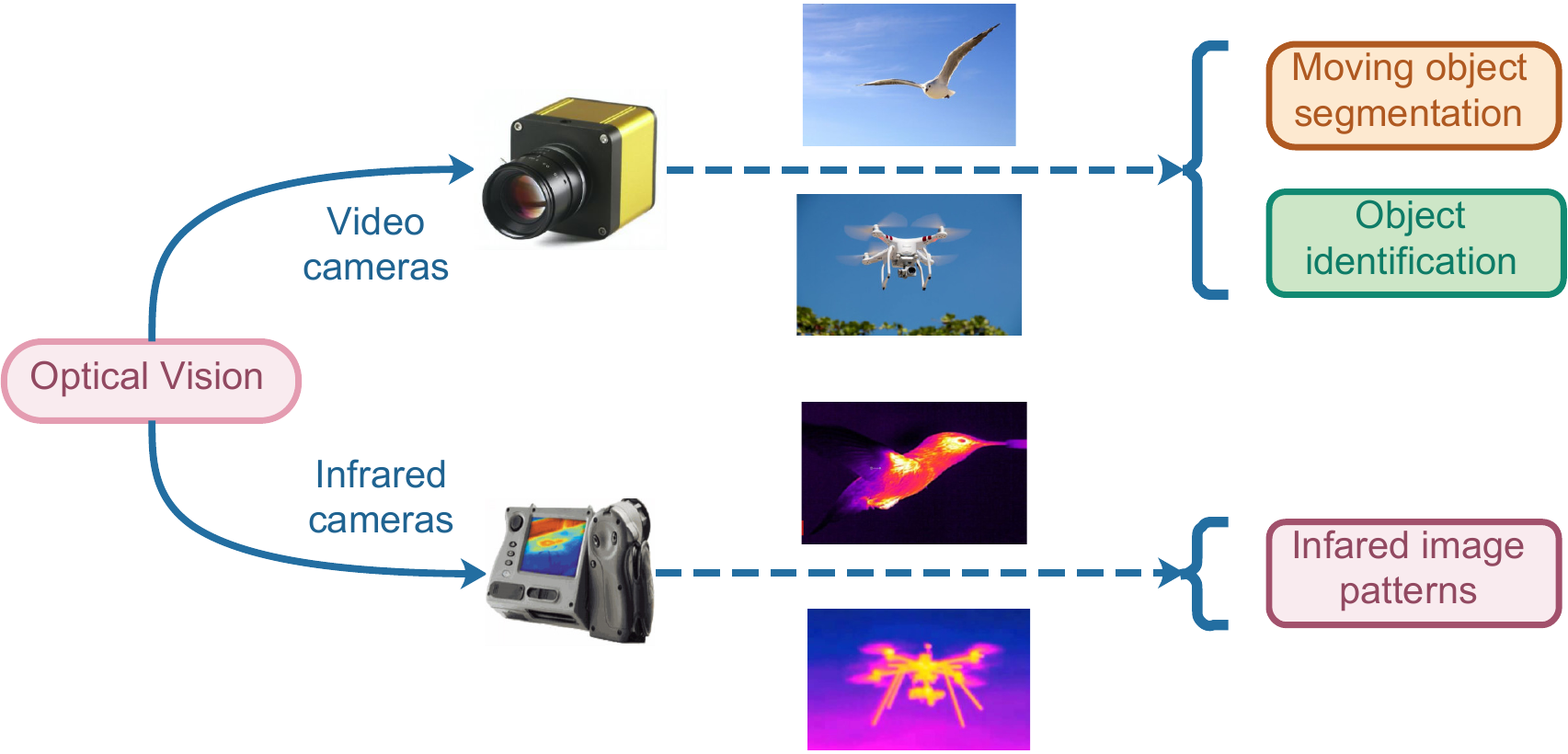}
    \caption{Categorization of vision based approaches}
    \label{figVisionDetectionCategories}
\end{figure}

Vision based UAS detection technologies mainly focus on image processing. Videos and cameras are adopted to capture the images of trespassing UAS. The ground stations figure out the appearance of UAS from the videos and pictures with computational methods. Conventional methods mainly rely on the methods of image segmentation. The differential of UAS and environment in images is used to determine whether the restricted areas have the UAS. A vision based UAS detection approach is presented in  \cite{V19}, which could separate the UAS from background efficiently. Similar work was reported in \cite{hengy2017multimodal} and \cite{christnacher2016optical}. The common challenges for their approaches are how to separate UAS from background images and how to distinguish UAS from flying birds. Typical vision-based UAS detection technologies are summarized in Figure \ref{figVisionDetectionCategories}.

In contrast, state-of-art image segmentation methods make use of neural networks to directly identify the appearance of UAS. An approach leverages the thermal camera to detect the UAS and neural network to identify the UAS \cite{V3}. An outstanding research is presented in \cite{V11}. A lightweight and fast algorithm which could operate on embedded system (Nvidia Jetson TX1) and identify the UAS in movement.

A real-time vision based UAS detection system is designed which is  based on two vision processing platforms: FPGA-based platform, which can operate below 10 Watt (i.e., power saving), and Graphics Processing Unit (GPU)-based platform, which is able to process more frames. However, for FPGA, it is impossible to change algorithms in real time \cite{V12}. Muhammad, S. et al. compared different convolutional neural networks' performance in detecting UAS, and their results showed that the Visual Geometry Group (VGG 16) network with Faster R-CNN achieved outstanding performance \cite{V13}. An approach is proposed to combine different pictures to generate synthetic images to extend the image data set to train the convolutional neural network to enhance the performance of the UAS detection \cite{V18}.

Birds are a serious factor which downgrades the identification of UAS from the images. Significant research efforts have been made to use convolutional neural networks to enhance the identification of UAS. Survey about the challenges of detection of UAS and birds is presented in \cite{V7}. The survey concluded that the neural network algorithms are promising in identifying UAS and birds. It compared policy based approaches and neural network based algorithms for the recognition of birds and UAS using datasets of videos and pictures. The results showed that the neural network based approaches can outperform the policy based approaches over 100 times in terms of accuracy and efficiency. An UAS detection framework is presented in  \cite{V9}, which is based on video streams and classified the objects into different types with convolutional neural network. The work mainly focused on distinguishing the birds and UAS in different scenarios. 

At the same time, some attempts were made to apply infrared cameras to identify the UAS. Infrared sensors  are leveraged to detect small variations of UAS in heat to identify the UAS. The drawback of this approach is that the heat from batteries has significant effects on result detection \cite{V15}. Different from other research on classifying the frame of images, dynamic vision sensors  are applied to capture the rotating frequency of the propeller to distinguish the UAS from birds efficiently \cite{V16}.

Currently, the vision based approaches can be implemented in some specific scenarios to recognize the features of UAS from the environment. The evolution of deep neural networks stimulated the processing of image processing which could have multiple positive effects on the UAS detection in vision field. The real time attempts showed that the vision approaches have the potentials of efficiency.  However, how to implement the recognized algorithms in multiple and variable environments is challenging. The novel approaches are supposed to be robust, adjustable, and precise. The vision based approaches need to be robust to the quick variation of the environment. The image distortion caused by weather change could be mitigated by the multiple level image processors which capture the features of UAS in different spectrum. The mobility of UAS poses a challenge to the vision based approaches, i.e., the images are supposed to be captured and recognized in different levels of mobility of UAS. The bio-inspired robots are limited in the detection accuracy, which poses the risk of distinguishing of UAS and birds mistakenly. The enhanced accuracy of the distinguishing can improve the efficiency of UAS detection and mitigation greatly.

\subsection{Radar based UAS Detection}
Radars have several advantages in detecting airborne objects compared with other sensors in terms of day and night operating capability, weather independency, and ability to measure range and velocity simultaneously. However, regular radar systems focus on fighting air targets of medium and large size with Radar Cross-Section (RCS) larger than 1 $m^2$, which makes it infeasible to detect small-size and low-speed UAS in  \cite{RADAR8384626} and \cite{RADARochodnicky2017drone}. It is difficult to detect UAS due to slow speed, since Doppler processing is typically used. Therefore, efforts are in needed to either develop new radar models or increase the detective resolution of conventional systems. In this section, we will  discuss three categories of radar based UAS detection technologies: Active detection, passive detection and posterior signal processing.
\begin{figure}
    \centering
    \includegraphics[width =0.95\linewidth]{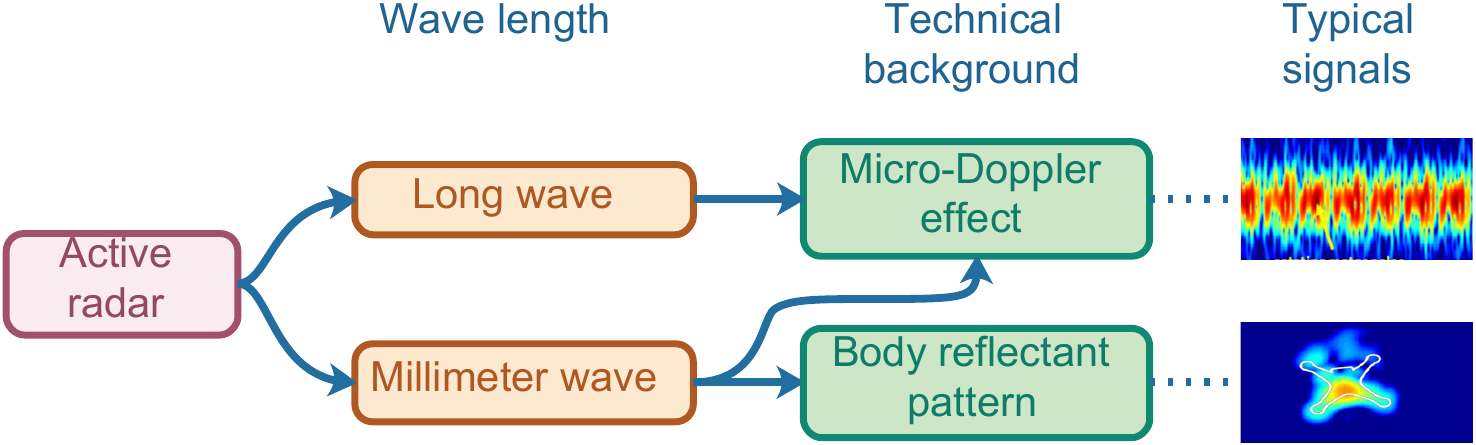}
    \caption{Categorization of active radar based approaches}
    \label{figPassiveRadar}
\end{figure}

\subsubsection{Active Detection}

Typically, there are two ways to increase the resolution of conventional radar detection systems for UAS surveillance: utilizing higher frequency carriers and using Multiple Input Multiple Output (MIMO) beamforming radio front-ends.

To utilize shorter wave length, in \cite{RADAR8228912} and \cite{RADAR8008143}, X-band and W-band Frequency Modulated Continuous Wave (FMCW) radars are designed for UAS detection. Their solutions use bi-static antenna and finally convert received signals into digital quadrature stream for posterior processing. The feasibility of using Ultra Wide Band (UWB) signals with 24GHz carrier was demonstrated in \cite{RADARnakamura2017characteristics}. The selection of carrier frequency for UAS detection radar should be higher than 6GHz (K-band) in  \cite{RADAR7153647} and \cite{RADAR7762208}.

Other approaches utilize multiple antennas to form MIMO front-ends. The benefit of such approach lies in its applicability to radar system with lower carrier frequencies. \cite{RADAR7590610} demonstrated the detection of a small hexacopter using 32 by 8 element L-Band receiver array, which achieved good detection sensitivity against micro-UAS. Similar research was presented in \cite{RADAR8447942}, a ubiquitous FMCW radar system working at 8.75 GHz (X-band) with PC-based signal processor is presented. The results indicated that it has the capability to detect a micro-UAS at a range of 2 km with an excellent range-speed association. In \cite{RADAR8008140} a Ka-band radar system which uses 16 transmit and 16 receive antennas to form 256 virtual antenna elements is presented. Their experiment proves that even in a non stationary clutter environment, the UAS could be clearly detected at a range of about 150m. Similar work was presented in \cite{RADAR7485236}. Another important characteristic of MIMO systems is that they generate large quantity of data for further processing in \cite{RADAR8378723}. The authors use the concept of data cube and classifier to assure the presence and location on an incoming UAS. A simplified approach, Multiple Input Single Output (MISO), was utilized for UAS detection in \cite{RADARsacco2018miso}. 

Noise radar is considered to be an efficient way to detect the slow moving UAS and its benefit is that UAS can be detected by using simple antenna components and lower carrier frequency. In \cite{RADAR8447988} and \cite{RADAR7799792}, the feasibility of using random sequence radar for UAS detection was demonstrated in sub X-band, and their results indicate that such radar can be the future of cost-efficient UAS detection solutions.

The advances in computation enable another radar, the SDR based multi-mode radar \cite{RADAR8059254}. Such radar is small-size and highly configurable. However, the operational performance of SDR relies highly on the back-end processor. In \cite{RADAR8405275}, two different implementations of FMCW radar and an implementation of continuous wave noise radar are presented to test their feasibility for UAS detection. And their findings indicate that the analog implementation has higher updating rate and Signal Noise Ratio (SNR). 

The apparent drawback of active radars is that they need specially designed transmitters which might not be easy to deploy and are vulnerable to anti-radioactive attacks.  

\subsubsection{Passive Radar}
\begin{figure}[H]
    \centering
    \includegraphics[width =0.95\linewidth]{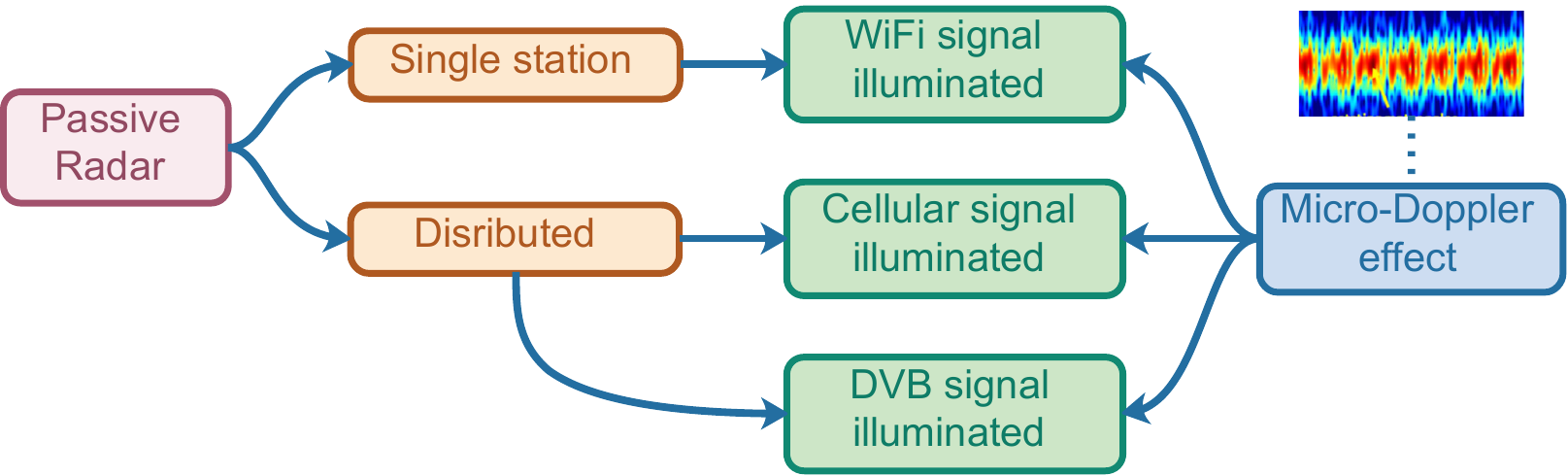}
    \caption{Categorization of passive radar based approaches}
    \label{figPassiveRadar2}
\end{figure}

Passive radars do not require specially designed transmitter. Existing radioactive sources such as cellular signals can be leveraged to illuminate the space. In this subsection, we classify passive radars into two categories: single station passive radar and distributed synthetic passive radar, as shown in Figure \ref{figPassiveRadar2}.

\begin{itemize}
    \item \textbf{Single station passive radar:}\\
    This type of passive radar exploits only one illumination source. The variation of received signals can be analyzed to specify the appearance of UAS. In \cite{RADAR8367508}, a WiFi based passive radar was presented for the detection and 2D localization of small aircraft. Obviously this is the most direct adaptation of active radars.
    \item \textbf {Distributed synthetic passive radar}\\
    Distributed station leverages the existing telecommunication infrastructures as illumination sources to enhance the UAS detection. There are mainly two approaches: cellular system based solutions and Digital Video Broadcasting (DVB) system based solutions.
    \begin{itemize}
        \item     \textbf{Cellular system based passive radars:} An approach is proposed to enhance the detection system which could locate and track the UAS by using reflected  Global System for Mobile communications (GSM) signal \cite{RADARknoedler2016detection}. An approach is proposed to receive the 3G cellular reflecting signal from UAS for track UAS. He leveraged the Doppler features of 3G cellular signal to monitor the target area and tracked the trajectory of UAS. The results showed that UAS could be tracked obviously in the waterfall data. The drawback of this approach is that it needs a reference receiver to calibrate the received signal. The accuracy of detection is dependent on the calibration accuracy seriously \cite{Ce3}. 5G mm-wave radar deployment infrastructure is constructed to detect amateur UAS. The deployed radars capture the signal of amateur UAS and upload it to cloud which could analyze whether there exist hazards. This approach could be an excellent approach in protecting safety of the city if the challenges of resource management, Non-Line Of Sight (NLOS) radar operation, noise mitigation and big data management could be addressed \cite{Ce2}. Similar implementation was presented in \cite{RADAR7735118}, where a passive radar array system is proposed to receive and process the Orthogonal Frequency Division Multiplexing (OFDM) echoes of UAS, which is originally transmitted by the nearby base stations. 

        \item \textbf{Digital video broadcasting system based passive radars:} The pervasive digital television signals are considered as an efficient illumination source for passive drone detection radars. In \cite{RADAR7944443}, \cite{RADAR8374028} and \cite{RADAR7944357}, passive drone detection radars were designed and tested. 
    \end{itemize}
\end{itemize}

Similar to active radar approaches, the micro-Doppler effect could be employed. In \cite{RFfu2018low}, UAS classification tests were conducted through propeller-driven micro Doppler signature and machine learning. Their experiment reveals that the micro Doppler signature of a plastic propeller is much less visible than carbon fiber propeller.

An apparent drawback for passive radar is that large amount of post processing efforts or multiple receivers are needed to achieve acceptable detection accuracy.

\subsubsection{Posterior Signal Processing}

\begin{figure}[H]
    \centering
    \includegraphics[width =0.95\linewidth]{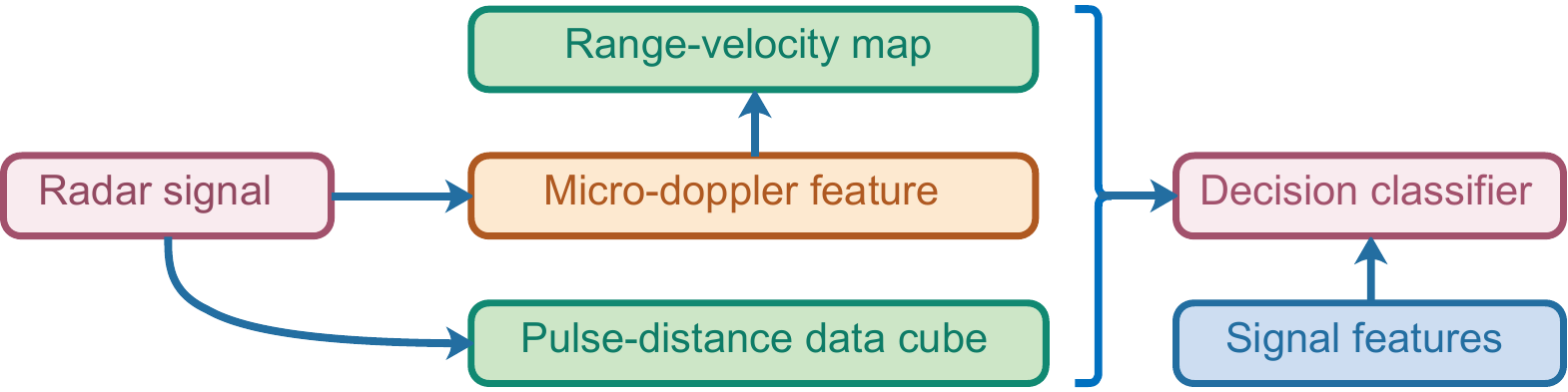}
    \caption{Categorization of radar signal processing.}
    \label{figRadarSignalProcessing}
\end{figure}

In UAS detection, efforts are needed to derive weak and sparse reflection signals of targets from noisy output of RF front-ends. Researches in this domain can be classified into two categories: conventional signal feature based detection and learning-based pattern recognition. The general categorization of radar signal is given in Figure \ref{figRadarSignalProcessing}.

   \begin{enumerate}
       \item \textbf{Signal feature based detection}\\
       The micro-Doppler effect of propellers of UAS are proved to be a useful feature of UAS detection. In \cite{RADAR8316996} and \cite{RADARjian2017experimental} methods for estimation of small-size UAS are discussed, with focus on the micro-Doppler signatures of rotating rotor blades. Their experiments proved that such feature can be used to distinguish UAS from other flying objects. In \cite{RADAR8059453},  a method based on hough transform was proposed to improve the detection and tracking performance. By making use of the linear distributed micro Doppler features, the method is able to detect and recognize UAS simultaneously. The micro-Doppler effect caused by flying UAS can also be used for detection of multiple UAS in \cite{RADAR8326374}, where the time\textendash frequency spectrogram is converted into the Cadence-Velocity Diagram (CVD). And then the Cadence Frequency Spectrum (CFS), as the basis of training data from each class, is extracted from CVD. K-means classifier is used to recognize the component of multiple micro\textendash UAS based on the CFS. Their experimental results on real radar data demonstrated that their method is capable of handling multiple UAS with satisfactory classification accuracy.
       \item \textbf{Learning based pattern recognition}\\
       Learning based pattern recognition methods is capable of classifying various types of objects. An example of conventional classification methods was presented in \cite{RADAR6875676}. This research demonstrated a classification experiment of UAS versus non-UAS tracks, based on a mixture of bird, aircraft and simulated UAS tracks, which is mainly based on the statistical features of the tracks and yields a high accuracy rate. Neural network based methods employ deep learning techniques to automate the feature engineering process of conventional machine learning. In \cite{RADAR8443549}, the UAS detection is carried by a CNN which learns the characteristic features in the 2D distribution of the Doppler spectrogram with high classification accuracy. In \cite{RADARmendis2016deep}, the Deep Belief Networks (DBN) are applied to characterize the features embodied by generated Spectral Correlation Functions (SCF) patterns to detect and identify different types of UAS automatically. The results of experiment illustrate that the proposed system is able to detect and classify micro UAS effectively. The validation of their approach using cognitive radio is given in \cite{RADAR8001610}. The benefit of learning based pattern recognition approach is that such systems are programmable and trainable to adapt to various scenarios.
\end{enumerate}
Radar based UAS detection can achieve better detection performance than the current other sensors. The antennas and signal processors were always considered as expensive options for the implementations. Though radar based approaches can achieve the better performance of the UAS detection (Long distance and Short distance), their cost is high in terms of deployment, calibration and maintenance. The manipulation of the radar based approaches requires the technicians to have the relative background of radar operations which poses a significant challenge to ubiquity in a large scale. The light, energy-saving, affordable and easy assembling radar element is desired which is supposed to be capable of easy deployment and maintenance. The posterior signal processing algorithms fueled by machine learning show great potentials to improve the efficiency and the accuracy of UAS detection. The improved capacity of reduced overhead of portable signal processing algorithms can expand the implementations of radar based UAS detection.

\subsection{Data-fusion-based UAS Detection}
Data fusion, which is the process of integrating multiple data sources to produce more consistent, accurate, and useful information than that provided by any individual data source, has the potential to generate fused data which is more informative and synthetic than the original inputs. The data fusion approaches can leverage the advantages of each method to acquire a combined result that is more robust, accurate and efficient than the single approaches. For the UAS detection, data fusion can be used to improve the performance of the UAS detection system to overcome the disadvantages of the single approach which exist in some specific scenarios.

\begin{figure}
    \centering
    \includegraphics[width =0.95\linewidth]{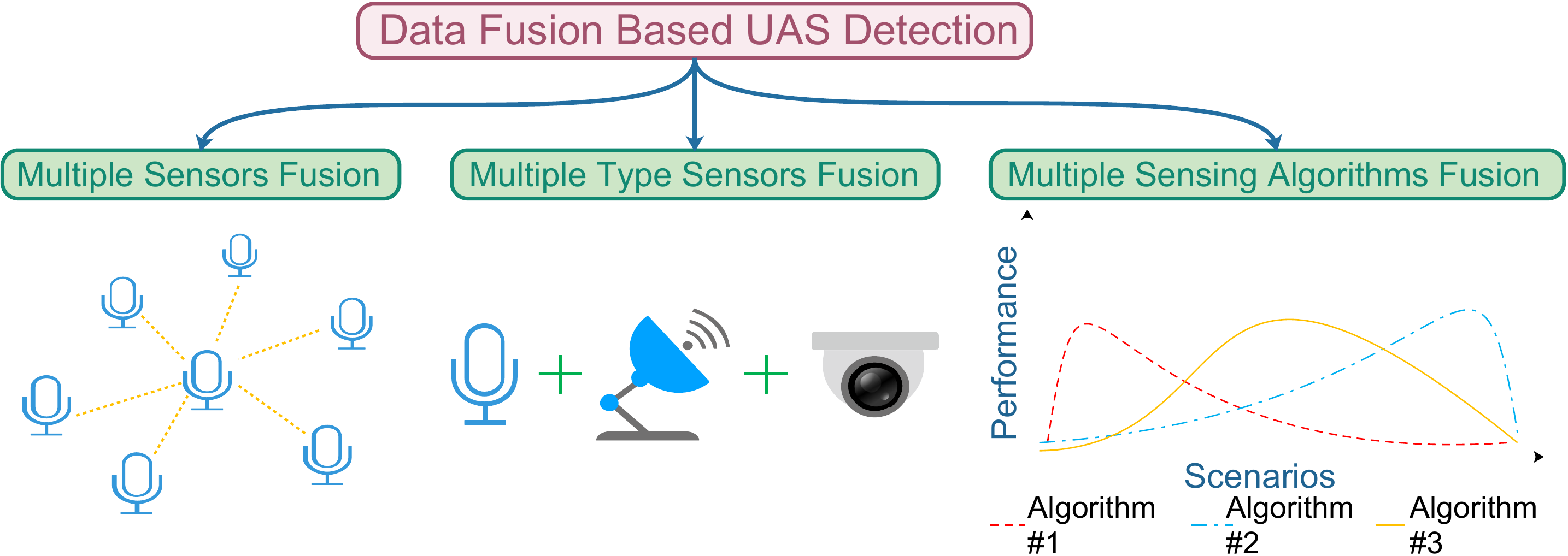}
    \caption{Data fusion based UAS detection}
    \label{fig_data_fusion}
\end{figure}

Based on the above discussion, the disadvantages and the advantages of each approach have been presented. The first general problem of the single approach is that limitation of the detection range and accuracy. The second general problem of the single approach is the vulnerability to detection scenarios. The third general problem of the single approach is the efficiency of the computation. Based on the above general problems, the research on the data fusion based UAS detection (shown as Fig.\ref{fig_data_fusion}), can be classified into three categories: 1) Multiple-Sensor Data Fusion; 2) Multiple-Type Sensor Data Fusion; 3) Multiple Sensing Algorithm Fusion.

\begin{enumerate}
    \item \textbf{Multiple-Sensor Data Fusion}\\
    Each type of sensors have their own advantages and disadvantages in the UAS detection scenarios. The general problem of the single approach is the limited detection range. The more straight and efficient approach to improve the performance of the single approach is designing a specific type of sensors to avoid the drawbacks of nature materials. \\ A classical example is acoustic sensor array. Distributed acoustic sensors are deployed in the detection areas. Each sensor can record the audio and deliver the record to the ground stations to make a combination evaluation of the environment in sound spectrum \cite{Uddin_2020}. The researchers extracted the phrase difference in the sound to locate the UAS. In the signal processing, the researchers need to adjust the weight of each sensor to obtain the higher accuracy of the location. Apart from the accuracy of detection, the data fusion can improve the functions of UAS detection. The RF based detection can acquire the RF signal with omni-directional antennas. The researchers could adjust the detection phrase of the signals in the receiving processing with multiple omni-directional antennas. Thereafter, the system with omni-directional antennas could acquire signals in the specific direction \cite{RF8337905,8008141}. This function could enable the ground stations to track trajectories of UAS and determine whether the UAS has the malicious intentions. The combined function also could enlarge the range of UAS detection which can not be gained by the single antenna.   The multiple sensor approaches combine multiple sensors with the same type to obtain the better accuracy or additional functions to improve the performance of the UAS detection.  The combination of the multiple sensors extends the capacity of the single sensor and maximize the range of detection geographically. 
    \item \textbf{Multiple-Type Sensor Data Fusion}\\
    In some scenarios, the improvement on the amount can not mitigate the disadvantages of the single sensors. Different UAS detection approaches are tested in  \cite{hengy2017multimodal}. The acoustic sensors are sensitive to the humidity, the temperature and the vibration in the environment. The cameras are invalid when the sunlight project on the lens directly. And the RF antennas are hard to recognize the target signal when it is buried in the white Gaussian noise environment.  More specifications can be found in
    \cite{hengy2017multimodal}. They concluded that the fusion of acoustic and radar could give more precise detection than other approaches. The conventional single sensors can not achieve outstanding performance on a variable environment, and meet multiple requirements of UAS detection ranges.\\ Concurrently, the cost of developing high quality functional sensors is prohibitively high. The researchers resort to the combination of different types of sensors. Based on the disadvantages and the advantages of each type of sensors, the researchers could combine different types of sensors to achieve the accuracy and long distance detection. Long and short range detection technologies are combined, where the passive RF receivers detect UAS's telemetry signals while video and acoustic sensors are used to increase the detection accuracy in the near field. Different range sensor systems including acoustic, optic and radars are integrated for UAS detection in target area. They deployed a 120-node acoustic array which use acoustic camera to locate and track the UAS, and 16 high revolution optical cameras to detect the UAS in the middle distance. In the long distance, they adopted MIMO radar to operate 3 different band radars to achieve remote detection \cite{Da2}. The resulting combination overcomes the drawbacks of each type of sensors on the UAS detection and maximize the advantages of each type of sensors. Simultaneously, the combination reduced the cost of deploying sensors in a large scale. In \cite{Da2}, the deployment of cameras can meet the middle distance detection requirement and reduce the cost on the deployment of acoustic sensors and radars. \\The combination of different types of sensors could achieve an outstanding performance for the restricted areas. However, the deployment and the configuration of this approach require much more specific and professional technologies and technicians with related backgrounds to maintenance. The different type sensor combination is a promising approach to maximize the capacity of sensors in the physical levels. In the future, investigation should be focused on exploring more combinations and characteristics of different sensors for more robust, affordable solutions.
    \item \textbf{Multiple Sensing Algorithm Fusion}\\
    The conventional approaches of data fusion fuse multiple data acquired from the sensors. Many data fusion approaches had maximized the capacity of sensors greatly. However, for UAS detection, the efficiency and the accuracy still can not meet the requirement. The novel approaches are needed to combine the multiple sensing algorithms to achieve the efficiency and the accuracy required by UAS detection. The sensing algorithms could be triggered according to the status of detection. The activated sensors deliver the information to the ground stations. Thereafter, the ground stations set the status of the detection, and the relevant algorithms will be swapped into the processing to extract the features of the target information. The UAS detection system will generate the threat outcome according to the result of sensing algorithms.\\ In this part, the sensing algorithms receive the data delivered from the sensors and extract the target features according to the types of sensors. The UAS detection system can adjust the sensing algorithm accuracy to meet the requirement of detection once the abnormal signals are detected. In \cite{Uddin_2020}, the researchers leverage the unsupervised approaches to extract the features of signal from various acoustic sensors under different scenarios (bird, airplanes, thunderstorm, rain, wind and UAS). Based on the recognition of scenarios, the system, proposed in this research, triggers support vector machine (SVM) and K Nearest Neighbor (KNN), separately, to detect the amateur drones in the restricted areas. To achieve the tracking efficiency, the authors, in \cite{10.1117/12.2532546, 8835545, 9002087, 9011293}, implemented multiple sensing algorithms on the passive mm-wave radar system to achieve the different accuracy of tracking according to the requirement of the recognition.\\ The combination of the multiple sensing algorithms could achieve advantages of the efficiency, the accuracy, less overhead of system and etc. according to the requirement of the UAS detection. However, how to make a reasonable arrangement for the sensing algorithms still needs more efforts. And the reasonable schedules of sensing algorithms can be a stimulation for the UAS detection system in the future.  
\end{enumerate}

Other data fusion schemes can be based on different platform integration. The researchers deploy multiple sensors into different platforms to leverage the mobility of different platforms, thus maximizing the sensors' capacities. The authors deployed the cameras on the surveillance UAS to make sure the amateur drones entering the restricted areas after the deployed acoustic sensors, in the sensing areas, send alarms to the ground stations \cite{10.1117/12.2304531}. In this research, they could recognize different targets such as birds. 


The data fusion approaches combine the advantages of each approach in the detection. The attempts show that the data fusion approaches have obvious advantages compared with single methods. According to the characteristics of each type of approaches, the detection deployment could contain multiple schemes in different areas which is apart from the center restricted areas in distance differently. Thereafter, how to implement the data fusion algorithms to achieve the consistency of the detection system on the results will be next challenge. Another challenge of the data fusion approaches is how to balance the weight of each approach in the final decision to achieve optimal detection results.

The distinction of \textquotedblleft capture and retrieve\textquotedblright and \textquotedblleft disable and drop\textquotedblright is important. Most malicious UAS are captured by the defenders with physical capture, Directional  EMP, RF jamming and Hacking. However, only the technologies of RF jamming and hacking can realize the function of retrieve. The retrieve function is supposed to be robust, accurate and efficient. The defenders are supposed to be confident that their systems have high probabilities of retrieving the malicious UAS again with protection of the property and the public. For the  \textquotedblleft disable and drop\textquotedblright, only the physical capture methods just drop the UAS from the flight. The technologies of Directional EMP, RF jamming and hacking have the both capacities of disabling and dropping. The Directional EMP, RF jamming and hacking can disable the UAS sensors, circuit, control system and communication devices to disable the control from remote attackers. However, these technologies can go deeper, like damaging control circuit and control algorithms, that can drop the UAS from the flight directly.

\section{State of the Art
Mitigation}
\begin{figure}[H]
    \centering
    \includegraphics[scale = 0.5]{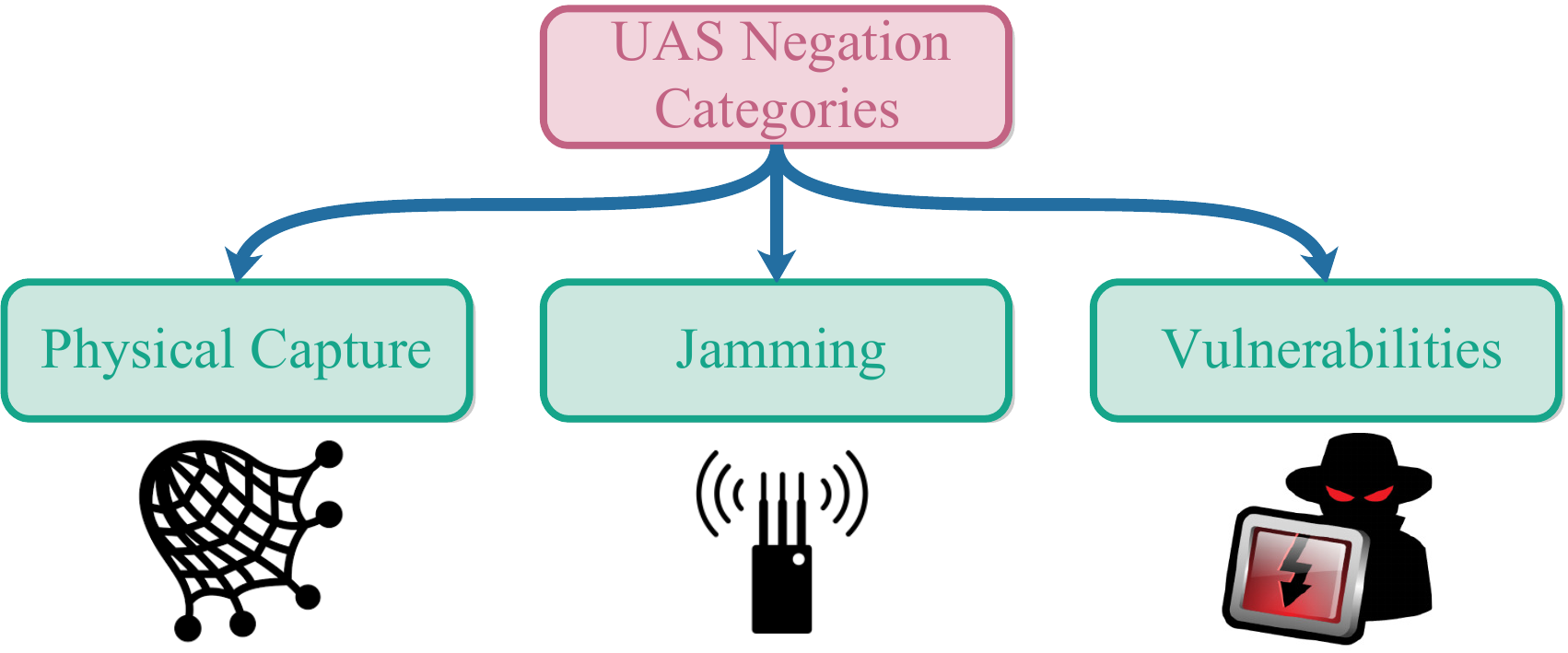}
    \caption{Categories of UAS Negation}
    \label{Archeteture}
\end{figure}

The technologies of detection and
mitigation are still immature. The research on UAS mitigation is limited. David etc. \cite{arteche2017drone} developed an architecture of UAS defense system. In this architecture, they specified the effective engagement range, including initial target range, detection range and neutralization range which is dominant for response. And their report showed that when the range is over 4,000 feet, the hardware based reaction and neutralization could operate efficiently. Based on the architecture, the approaches could be classified into three main categories, as shown in Fig.~\ref{Archeteture}. The first one is the physical capture which focuses on capturing UAS with physical methods. The second is to leverage the noise generator to jam the systems or sensors, thus rendering the UAS inoperable by the UAS controller. The third is to exploit vulnerabilities of system or sensors to acquire priority of control.
\subsection{Physical Capture}
\begin{enumerate}
    \item Nets capture\\
    Net capture is a physical method to negate the UAS. The defenders adopt guns or some specific weapons to trigger the net to catch UAS. The net is stretched when the net is shot, and closed to disable the mobility of drone when the net touches the drone. Kilian etc. \cite{kilian2015counter} invented a deployable net capture system which could be installed in the airplane or authenticated UAS. When the unauthorized or unsafe UAS are located, the system could capture the unauthorized or unsafe UAS. Practical approaches to neutralize UAS are attracting attentions of the military. In \cite{Tomase2019Scalable}, a spin launched UAS projectile is developed. This projectile aims to launch a net to capture a flying UAS. The net is stored in the warhead of projectile  which allow soldiers to shoot it by regular guns.
    \item Directional Electromagnetic Pulse\\
    Electromagnetic pulses have been mainly used to counter illegal electronic facilities in the car which could restart or disable the operation of control system. Based on the function of electromagnetic pulse, Gomozov etc. \cite{8100595} focus on the functional neutralization of on-board radio electronic system on UAS, and they adopted spatio-temporal pulse of the wavelength $\lambda = 2.5cm $ to neutralize the UAS and their results showed that their approach could provide an aimed impact on UAS in the range from 0.5 to 1 km and no harm to biological protection.
\end{enumerate}
The physical capture mainly focuses on disabling the mobility of drone and control system. The physical capturing approaches have advantages of easy manipulation, light weight, quick assembling, etc. Once the drone is captured by the physical capturing approaches, the drone will experience damages at different levels. The physical capturing approaches are efficient and low cost, but not friendly to pilots. 
\subsection{Jamming}
\begin{figure}[t]
    \centering
    \includegraphics[scale = 0.28]{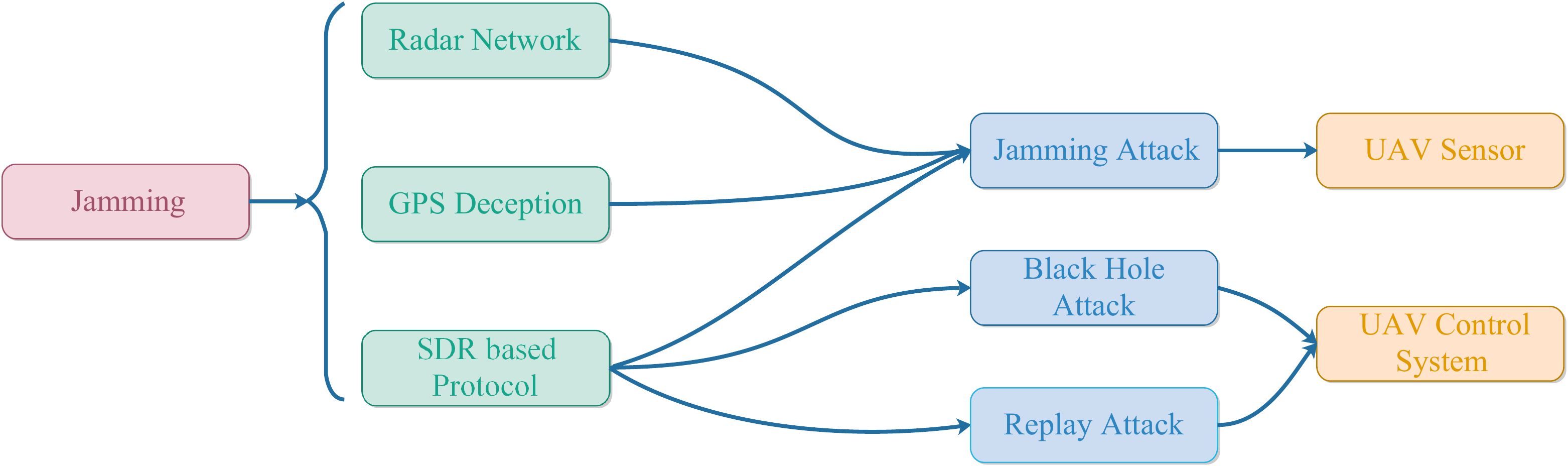}
    \caption{Categorization of Jamming}
    \label{jamming}
    \end{figure}

Jamming is the most popular method used in neutralizing UAS entering restricted areas. The defenders leverage noise signal to interfere operation of UAS sensors or systems for neutralization. In this subsection, we classify jamming into three main categories, as shown in Fig.~\ref{jamming}. Among these attack methods, the main targets are UAS sensors and systems. Zhao etc. \cite{5367397} proposed an approach to leverage a team of UAS to form an air defense radar network which could jam the targets' sensors. This approach could detect and negate the unauthenticated UAS and their experimental results showed that they could track and jam the phantom made by DJI to leave the restrict areas, and proofed that the N UAS, in a team, could negate at most $N\times (N-1)$ targets. Li etc. used the direct track deception and fusion to invade the control priority of navigation system and trajectory control system. Based on the GPS deception jamming theory, they leveraged the trajectory cheating to lead the unauthenticated UAS to fly out from the restricted areas. Their results showed that both the direct track and the fusion track could make UAS drift off the restricted areas. P\"{a}rlin etc. \cite{8398711} proposed an approach to use the SDR to realize a protocol-aware UAS jamming system. They used an SDR to achieve remote controller's signal and recognized the communication protocol with analysis. The SDR generates command to control the UAS to fly away restricted areas. They compared three different approaches (Tone, Sweep and Protocol-Aware) to evaluate the performance of their approach which showed that the protocol-aware is more efficient than tone and sweep jamming. 
\begin{itemize}
    \item \textbf{Tone:} a narrow band signal at the center of a single channel.\label{tone}
    \item \textbf{Sweep:} a linear chirp swept across the entire 2.4 GHz ISM band.\label{sweep}
    \item \textbf{Protocol-Aware:} a signal imitating either Futaba Advanced Spectrum Spread Technology (FASST) and Advanced Continuous Channel Shifting Technology (ACCST).\label{protocol}
\end{itemize}
Li etc. \cite{8438310} proposed an approach to leverage the UAS with jammer to neutralize other UAS eavesdroppers. They used the mobility of UAS to get close to malicious UAS and the jammer installed on the UAs could impact the malicious UAS' trajectories. Sliti etc. \cite{8473921} presented different attacks which focus on UAS network for neutralization.
\begin{itemize}
    \item Jamming attack: Transmitting a jamming signal to disrupt communications between a drone and the pilot, forcing the drone to return to “home” location, i.e., where it took off.
    \item Black hole attack: a type of denial-of-service attack which discards the incoming or outgoing traffic of communication.
    \item Replay attack: a network attack that vises to maliciously repeat a valid communication so that the communication of the UAS could be analyzed and invaded. 
\end{itemize}
Curpen etc. \cite{du123} focus on neutralizing the UAS which is based on Long Term Evolution (LTE) network. The spectrum analysis in two different cell networks showed that the efficient jamming range for LTE UAS is approximated 60m. Mototolea etc. \cite{8484821} leveraged the SDR to analyze and hijack the small UAS. In this work, the decoding protocol of DSM2 could get the fingerprinting and pairing process. Willner \cite{willner} invented a system which could neutralize remotely explosive UAS in a combat zone. This system could be equipped on the ground or installed on the authenticated UAS which transmit the jamming signal once the target is locked. Bhattacharya etc. \cite{5530755} developed a game theoretic approach to optimize the jamming method on the expelling the UAS attacker.

Jamming approaches can provide friendly and zero-damage schemes to neutralize the drone entering the restricted areas. The jamming approaches can provide the neutralizing effects on the drones in different levels (from hardware to software). The jamming can be deployed in a large scale and take effect for a long time. However, the current jamming can not make a directional effect which can be controlled by the defenders. The effects of jamming is omni-directional which could affect the devices in the restricted areas, and energy consuming. The jamming needs a long time to take effects when the drone receives enough jamming signals. In the coming future, the jamming is supposed to be controlled, directional and quickly reactions.
 
\subsection{Vulnerabilities}
\begin{figure}
    \centering
    \includegraphics[scale = 0.45]{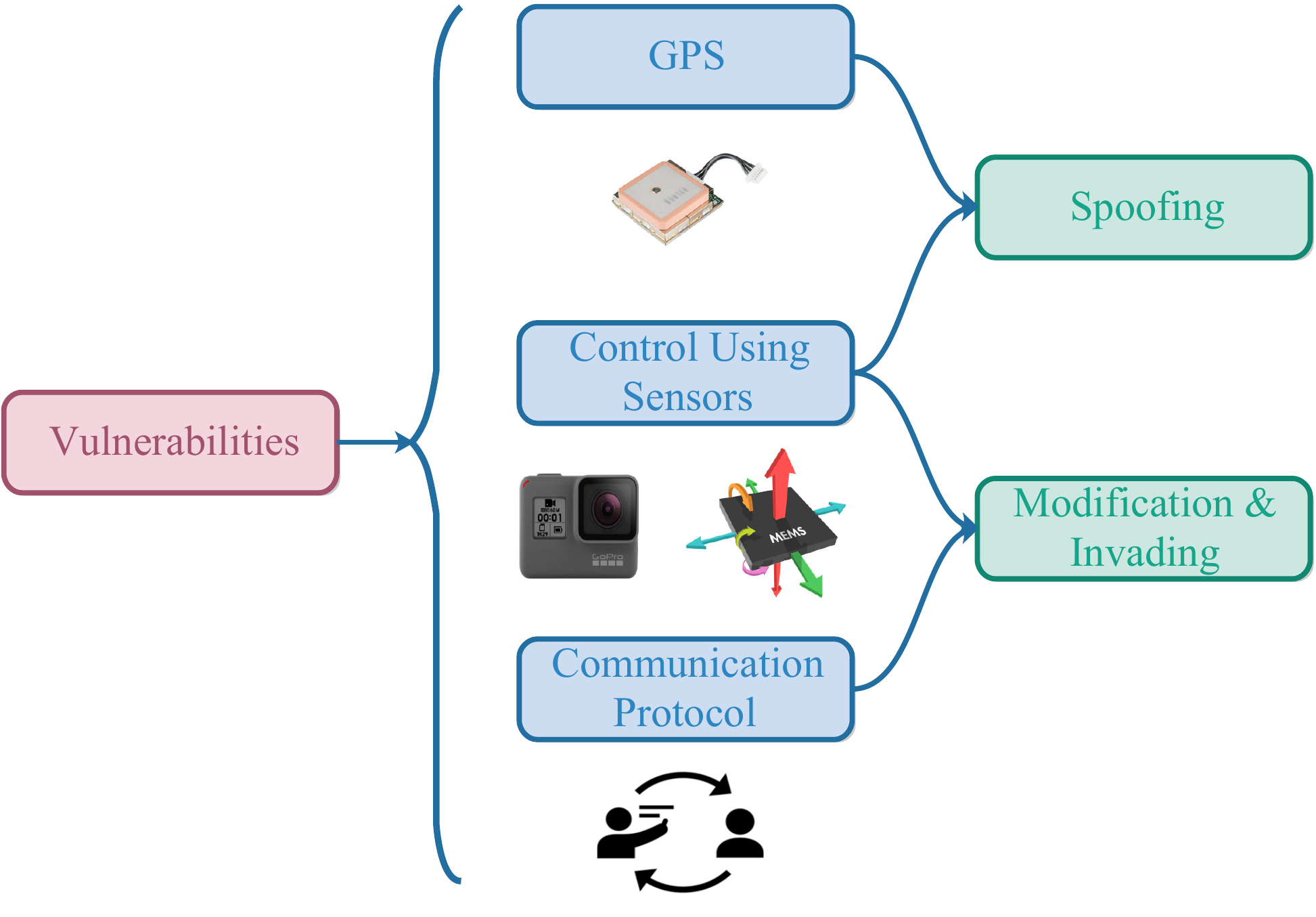}
    \caption{Catergorization of Vulnerabilities}
    \label{Vulnerabilities}
\end{figure}

There are three main methods to exploit the UAS vulnerabilities, as shown in Fig~\ref{Vulnerabilities}. Most vulnerabilities exploitation work focus on GPS control using sensors and communication protocol. The defenders leverage the spoofing methods to GPS and control using sensors, and adopt modification and invading to control using sensors and communication protocol. Rodday etc. \cite{7502939} demonstrated an approach to exploit the identified vulnerabilities of the UAS control systems and performed Man-in-the-Middle attack to inject the control commands to interact with the UAS. Dey etc. \cite{8326960} presented cracking SDK, reversing engineering and GPS spoofing to hijack the UAS. They compared the DJI and Parrot drone performances under the exploitation attack. The results showed that the DJI is more secure than Parrot, which means that the DJI is hard to rush into the restricted area. Chen etc. \cite{8406948} analyzed the popular altitude estimation algorithms utilized in navigation system of UAS and proposed several effective attacks ( shown as \ref{Item1} ) to the vulnerabilities. 
\begin{itemize}
\label{Item1}
    \item KF-based Sensor Fusion
    \begin{enumerate}
        \item Maximum false data injection: modifying the calculation functions of GPS and barometer measurements. 
    \end{enumerate}
    \item Altitude estimation based on accurate sensor noise models
    \begin{enumerate}
        \item Blocking GPS: Disable the GPS readings.
        \item Modifying barometer readings: Manipulating the barometer and inject bad data.
    \end{enumerate}
    \item First-order Low-pass filter: Inject the barometer readings and influence the accuracy of estimation. 
\end{itemize}
Esteves etc. \cite{8484990} demonstrated a simulation on a locked target to gain access to the internal sensors for neutralizing the UAS from restricted areas. Melamed etc. filed a patent on how to utilize the SDR with antenna array to detect the UAS and create override signal to link of communication to neutralize the UAS. Marty \cite{marty2013vulnerability} presented an approach to hijack the MAVLink protocol on the ArduPilot Mega 2.5 autopilot. Katewa etc. \cite{8062641} proposed a probabilistic attack model to neutralize the UAS which executed denial of service attack against a subset of sensors based on Bernoulli process. They also described the vulnerabilities on the sensors of the UAS and strategies to negate the UAS via jamming sensors or access control systems. Huang etc. \cite{8350330} proposed a spoofing attack based on the physical layer to utilize the angle of arrival, distance-based path loss, and the $Rician-\kappa$ factor to recognize the UAS and source where the signal comes from.

Penetrating vulnerabilities of system has affected the processing of security in computer field for a long time. The evolution of the amateur drones enables the drones have their own Operation System (OS) which gives the defenders a great chance to exploit the system via the vulnerabilities of OS. The integration of embedded system and sensors extends the vulnerabilities of the OS of drones. The vulnerabilities releasing of the OS for the drones could improve the success of exploit the unauthorized drones with malicious intentions in the future.

For the deployment of detection technologies, the deployment decides the capacity of each type of approach significantly. According to the nature of sensors, the deployment on ground based stations, UAS or manned aircraft are varying. The acoustic sensors are sensitive to the sound which needs the environment noise keeps stable and quiet. This means the acoustic sensors are not suitable for the mobile platforms. The passive RF based detection has specific requirement of the antennas distance between each other which is important to achieve accurate result for passive RF signal detection. The passive RF based detection needs the platform have powerful computation capacity which are just for the ground based stations and manned aircraft. The current UAS can not provide suitable computation and power supply. The vision based detection can be deployed on the ground based stations, UAS and manned aircraft. The key sensor of vision based detection are mainly cameras. Meanwhile, a number of light weight, low energy consumption cameras are suitable for the deployment of vision based detection. The current radar based detection has the disadvantages of heavy weights which is a challenge for the mobile UAS platform. The payload  and power supply of the UAS are limited which can not satisfy the requirement of radar. The most cases of the deployment of radar based UAS detection are ground based stations and manned aircraft. The most fancy approaches are the data fusion based detection which are just constrained by computation capacity. Concurrently, the data fusion based detection approaches are mainly implemented in the ground based stations and manned aircraft. Apart from the above, the combination of different platforms also has promising potentials to improve the capacity of detection for big properties. The different deployment on the ground based stations, UAS and manned aircraft could achieve scalable detection to malicious UAS.

\section{Challenges in UAS Detection and Mitigation}
Tables IV and V compare various UAS detection and mitigation technologies, respectively. On one hand, millimeter wave radar along with data fusion methods are considered as the most promising trends for UAS detection in the future, on the other hand, physical capture is regarded as the most practical and reliable approach to neutralize unwelcome UAS. Hacking and spoofing have emerged as a promising negation solution with low footprint and low collateral damage. However, there are a lot of challenges which must be addressed to develop mature scalable, modular, and affordable approaches to UAS detection and negation. In this section, we will identify the challenges of each UAS detection or negation technology.
\begin{figure*}
    \centering
    \caption*{TABLE IV: Comparison of UAS Detection technologies}    
    \label{figMethodComp}

    \includegraphics[width = \linewidth]{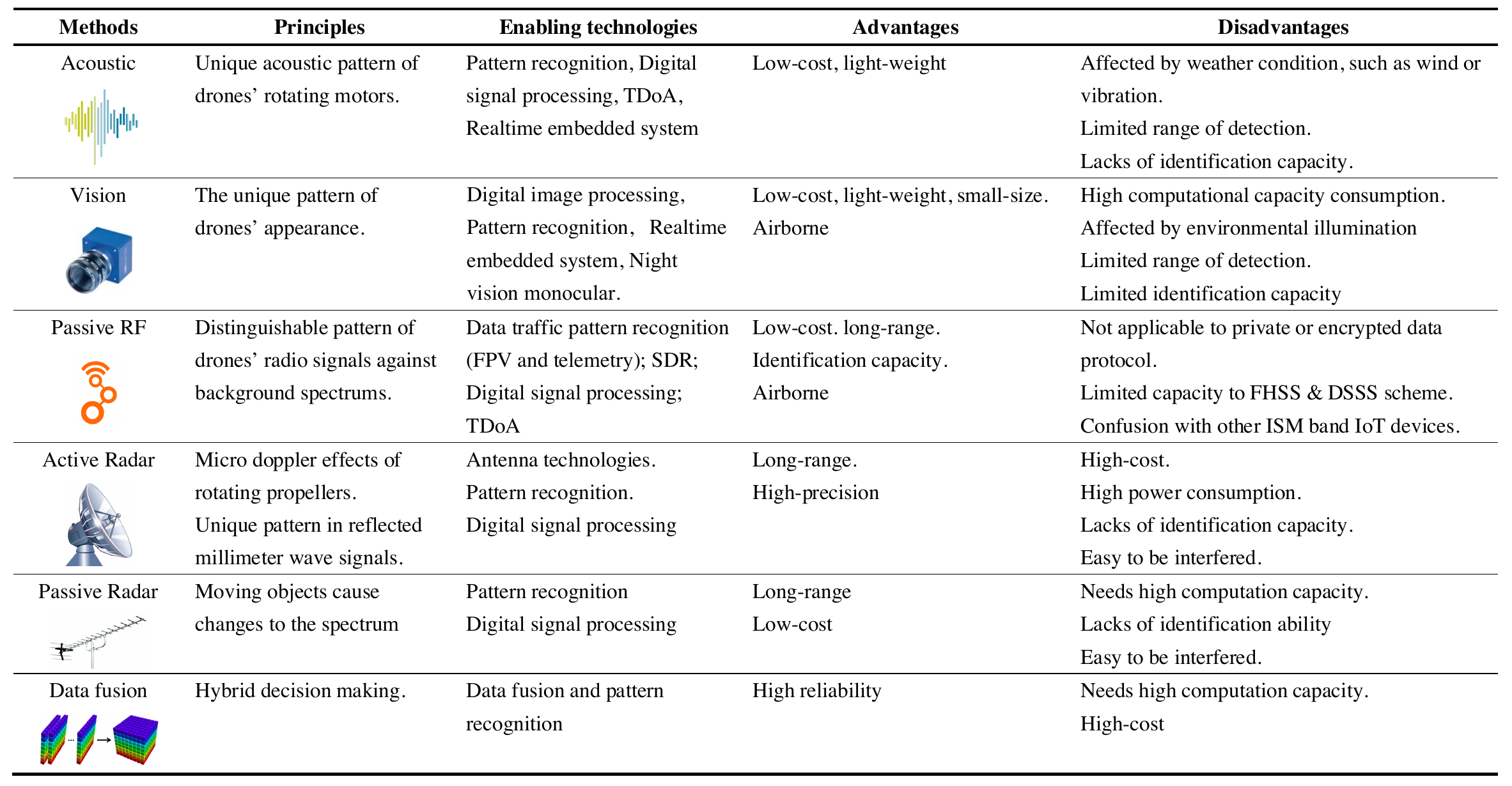}
    
\end{figure*}
\subsection{UAS Detection}
\subsubsection{Acoustic based UAS Detection}
The piezoelectric substrate materials are the core of the acoustic sensors. These materials could generate the electricity according to the strength of vibration on the surface which is very smart characteristics. But these materials also could be affected by the temperature, humidity and light intensity. In practice, it is hard to keep a stable performance of the detection when the scenario is full of multiple variable physical parameters. At the same time, the acoustic sensors are sensitive to the vibration of the air caused by wind. The real UAS signal with the multiple fading effect may be buried in the wind. The acoustic sensor array can improve the accuracy of detection, and multiple sensors in different places could locate the position of the UAS, but there is limited research which takes into account Doppler effect generated by the movement of the UAS, and the combination effect of the movement of wind when the speed of wind is over $5~m/s$, because the effect caused by wind is not negligible.

\subsubsection{Passive RF based UAS Detection}
The passive RF needs multiple antennas to form an antenna array and recognize the UAS according to the combination of each antenna's detection results. The passive RF detection methods are highly dependent on the telemetry protocol and RF front-ends. A novel approach which could recognize multiple protocols simultaneously are needed. And such an approach is supposed to be efficient, stable and easily deployed in mobile embedded devices. To our knowledge, there are no SDR specifically designed for UAS detection available on the market. Currently SDR devices suffer from heavy weight, high energy consumption, and poor potability, which limits the use of SDR in UAS detection. 
How to design a light-weight, small-size, and low cost RF analysis device with comparable or better performance could be a significant improvement for the UAS detection. At last, the artificial intelligence (AI), like deep learning, could be a good approach to improve accuracy and robustness of the RF based detection methods. With the plenty of signal data generated per second feeding, the deep learning has the potential to improve the accuracy and efficiency. The emergency of the RF signal in different small scales is very important for efficient and accurate recognition for UAS.                
\subsubsection{Vision based UAS Detection}
Although vision detection has been investigated for a long time, research efforts are still needed to improve their performance. Firstly, there is a urgent need for a vision device designed for UAS detection to be light-weight, small-size, and low cost. The most important issue is that how to adjust the aperture of the camera to avoid the fading effect caused by sunlight in different angles. Secondly, the shape of birds is very similar to some fixed wing UAS, and many bionic robots could fly like birds. It is urgent to classify these two scenarios in the mobile devices, especially these devices could be installed in the surveillance UAS. So combining some additional biological signals in the detection process is needed to recognize these two different situations. Thirdly, the deep learning has been applied in vision detection for many years, but low size, weight and power-consumption (SWaP) deep learning algorithms are still needed. The portable deep learning algorithms could be implemented into a new scenario without too much time training. In the computation field, this function of deep leaning is called transfer learning. The mature and outstanding models of deep learning can be implemented to multiple scenarios to achieve excellent performances with few training episodes.  

\subsubsection{Radar based UAS Detection}
There have been significant research efforts made on the radar based UAS detection. The grounded radar could well meet the requirement of UAS detection in military. But for the civilian usage of the UAS detection in the scenarios like stadiums and residential areas, the current schemes are not easy and fast to be deployed in the places with the crowds. Most radars are highly dependent on the antennas which are very huge and lack of flexibility. The light-weight, small-size and mobile based phrase array radars for UAS usage have potentials to meet the requirement of deploying in the civilian scenarios. This radar also needs to be low-cost, because UAS does not allow the energy consuming equipment to be on board. Another challenge for the radar is how to solve the interference between the radar and other communication equipment on the UAS.
Based on advanced radar detection equipment, there also needs an optimized deployment approach to change the radar deployment according to the change of scenarios in real time. Of course, the fast and accurate radar analysis approaches also can make a great improvement of the detection of the UAS. 

\subsubsection{Data Fusion based UAS Detection}
Multiple signals fuse from different detection devices which have different data format. The traditional methods perform well on the different data in the same acquisition accuracy, but can not fuse well in different acquisitions. To detect unauthorized or unsafe UAS, the surveillance needs to fuse multiple data from different devices including video, radio, and audio, and so on. A novel fusion approach could input the data in different formats simultaneously and easily port in different embedded systems. 

\subsection{UAS Mitigation}
\begin{figure*}
    \centering
    \caption*{TABLE V: Comparison of UAS Mitigation Technologies}    
    \includegraphics[width = \linewidth]{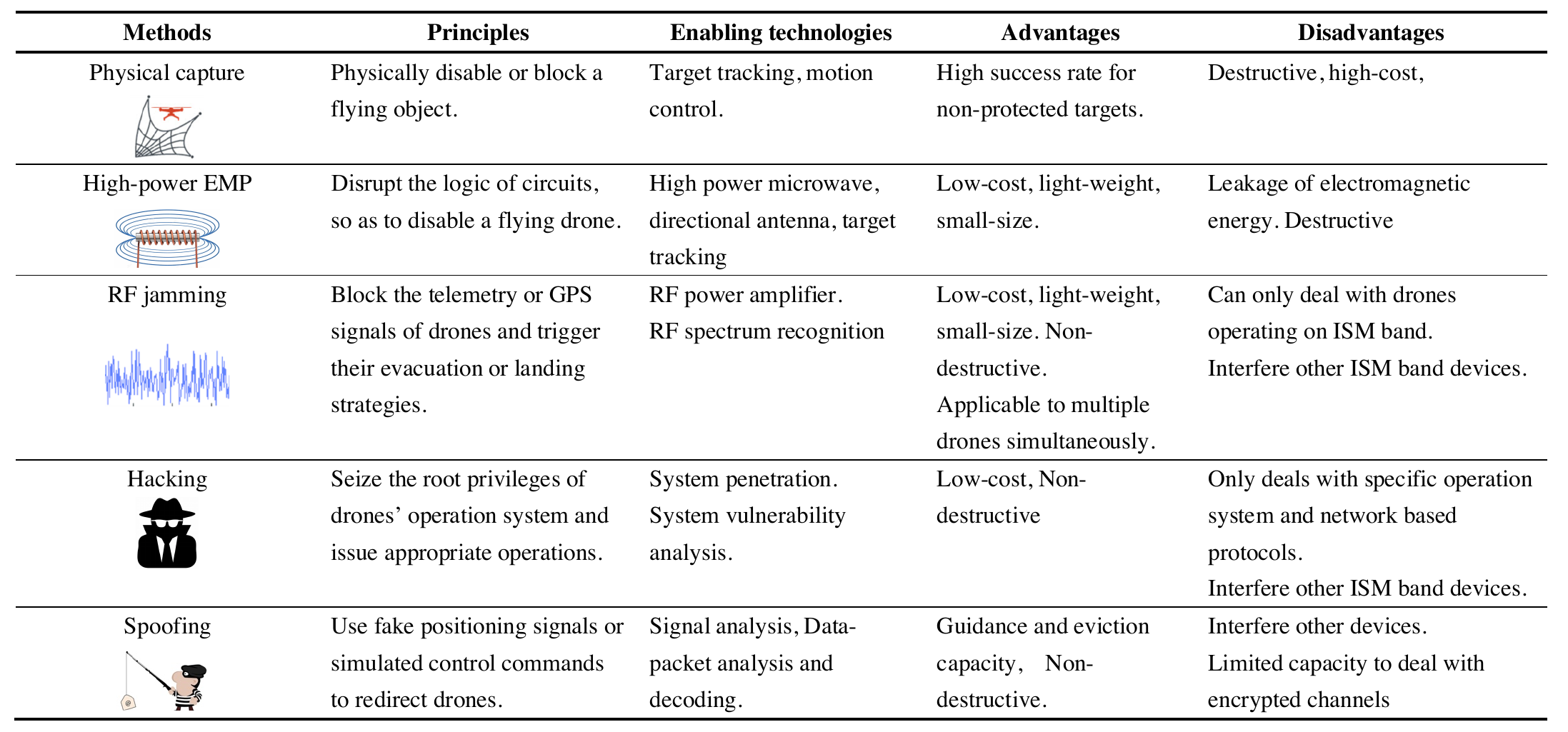}
    \label{figMethodComp2}
\end{figure*}

\subsubsection{Physical Capture}

The physical capture is the most direct way to counter unauthorized or unsafe UAS, which is easy to be deployed. So the requirement of the physical capture methods is light-weight, variable scale, and easy to master. 
\begin{itemize}
    \item The physical capture needs to be light so that the human being could carry it and get ready to take down the UAS once they make sure that the UAS is unauthorized or unsafe. Also, the surveillance UAS could install the equipment and the surveillance UAS could counter the intrusion UAS when they are patrolling.
    \item The physical capture needs to be variable scale so that it could work when there are many different styles of UAS in size. In the current, the most effect approach is net, but the net also needs to be optimized which should be light, firm and recyclable. The benefit of the net is that it could capture all the things in the capacity of net. Apart from the net, the bolas also works well when capture the UAS. The bolas is made of weights on the ends of interconnected cords which is efficient when the target has frames or propellers. The power system of the most UAS is based on the propeller. The bolas could be a powerful tool to stop the UAS working if the rope is strong enough. The net and bolas both need to be designed in variable scale so that they could be used to capture UAS in different sizes.
    \item The physical capture should be easy to master. The physical capture tools like nets and bolas, could be loaded into an easy trigger platform like bullets so that a proper shooting gun could trigger the bullets to the target and capture UAS. The surveillance UAS just carry a light and simple trigger platform without much energy consumption. To improve the successful capturing rate, the physical capture methods needs assistance of the navigation and tracking systems like missiles. 
\end{itemize} 

\subsubsection{Directional EMP}

The directional EMP is a very efficient weapon to counter the UAS which could navigate itself by Inertial Measurement Unit (IMU) without any communication with outer facilities. The main challenges of the directional EMP are no harm to human being, low divergence angle and long distance effect.
\begin{itemize}
    \item Most EMP shooting contains too much electromagnetic energy so that it also has damage effect on the other nearby facilities or human beings. To make the EMP shooting approaches more practical, the researchers need to find a proper frequency that take efficient effect on UAS and zero damage to organism. Because the organism obtains the electromagnetic materials, if the EMP frequency is similar to the respond frequency to the humans or animals, the organism will take response to the specific frequencies. We need to make sure that the EMP weapons work in a different frequency from human and animal response.  
    \item The EMP is high power weapon which needs much electricity to drive the EMP. However, the divergence angle of the EMP wave will cause the much fading of energy when it has effects on the target. So the research needs to reconsider how to minimize the divergence angle of the wave. The narrow divergence angle of the wave could improve the success rate of countering UAS while saving power.
    \item There is a disadvantage of the EMP.  Different frequencies have different transmission distances. The higher frequency EMP will disappear more quickly in the air while the higher frequency EMP obtains more energy which improves the counter successful rate. So there is an embarrassing problem: when the target is detected but the counter distance is limited. How to balance the frequency and distance to achieve a better performance of the counter purpose is desired.
\end{itemize}
\subsubsection{RF Jamming}
 The current RF Jamming approaches have the following disadvantages. First, the RF jamming consumes too much power which is not practical for surveillance UAS to carry to execute the RF jamming precisely. Second, RF jamming works for an area, however the RF jamming could not jam a specific target in a desired point. The RF jamming only takes effect when the UAS are communicating with outer devices. The RF jamming does not work if the UAS navigates itself with inner global navigation system (GPS). 
 \begin{itemize}
     \item The current RF jamming devices are deployed on the ground or mobile vehicles which have limited movement space to generate the efficient jamming signals to interfere the UAS communication system. Also these devices are too heavy to be loaded on the surveillance UAS. So the light-weight, high efficient and low cost RF jamming devices are required, especially the UAS oriented RF jamming devices.
     \item RF jamming has similar characteristics to electromagnetic wave. So once the target is determined, how to transmit the jamming signal to a specific position is an issue. The directional antenna could send the electromagnetic wave into the specific area, but it still needs to realize the specific points attack. 
     The phase array radar maybe a good direction for further investigation. A RF jamming array may include the omni-directional antennas and directional antennas. The control ends change the RF jamming transmitting power and phase to realize a combination of RF jamming in a specific position.
     \item The RF jamming is easy to realize but how to recognize the target communication channels is also an open problem. The surveillance officials could eavesdrop the target UAS communication and determine its communication channels. With the recognition of communication, the defenders just execute the RF jamming in the specific channels and the energy on the invalid jamming channels will be saved.
 \end{itemize}
 \begin{figure*}
    \centering
    \includegraphics[width = 0.7\linewidth]{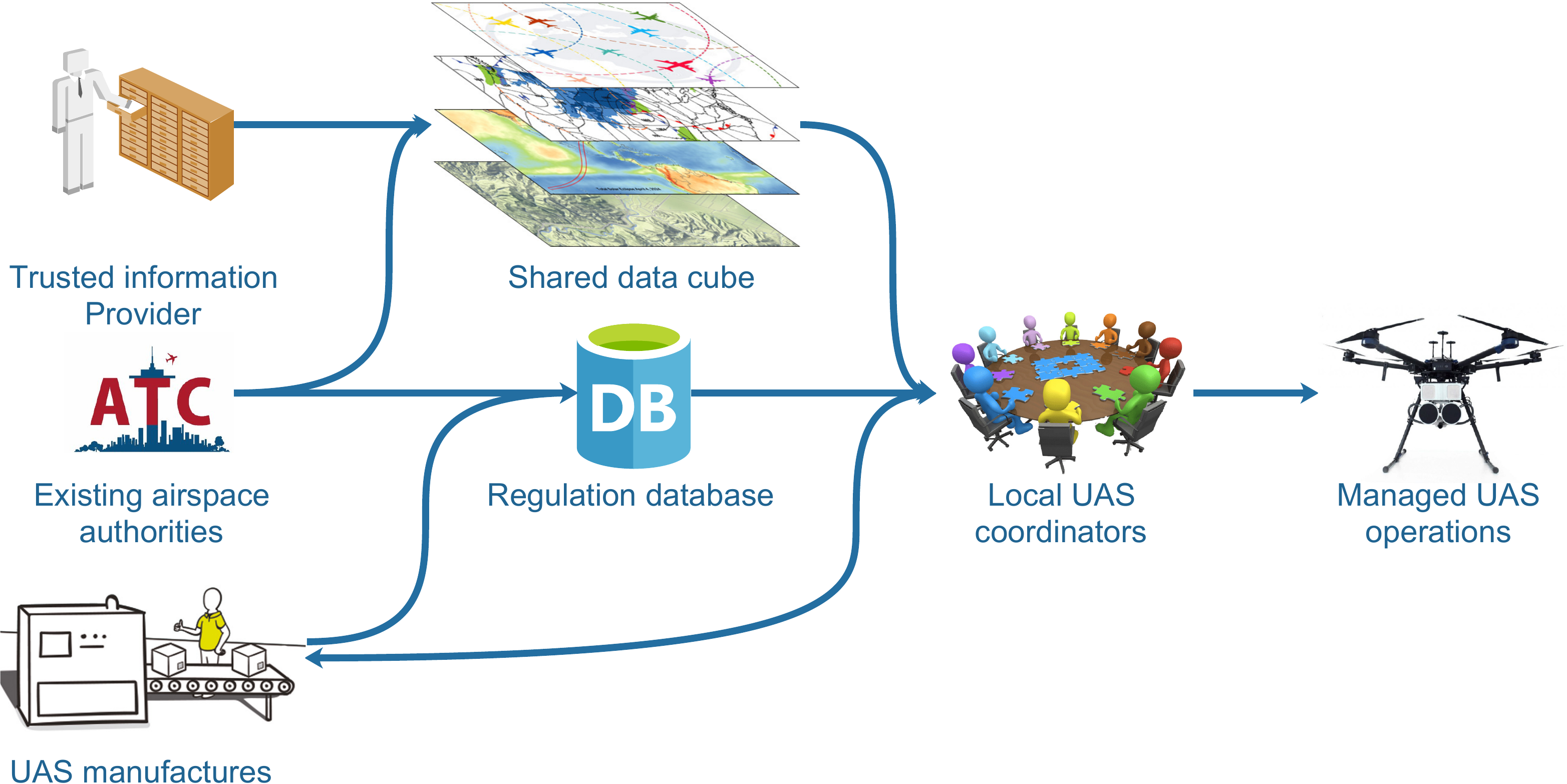}
    \caption{Unified Framework for Drone Safety Management}
    \label{figUnifiedFramework}
\end{figure*}
\subsubsection{Hacking}
Hacking UAS has been investigated for many years. The hacking methods mainly focus on outer interference, networking, and spoofing.
\begin{itemize}
    \item The current outer interference requires that the interference devices are very close to the UAS sensors (like IMU, GPS) so that the UAS could get the correct data from the sensors. The remote interference devices and approaches are needed to lead the UAS to fly away from the sensitive area in a long distance. 
    \item The network using on UAS are mainly WiFi and cellular networks. One way is to attack the WiFi and cellular networks of the UAS to obtain the priority of the UAS, and send command to autopilot to lead the UAS to launch off in a safe place. Also, the defenders could leverage network to access the link and play the man-in-the-middle, decrypt the communication packet of the command, and then modify the command to control the target to launch or fly back where it comes. There are also another approach to allow the cellular operators to request authentication on the cellular link to check the commands on the flight randomly.   
    \item There are some other sensors that the researches have not tried yet like optical flow, camera and laser. The optical flow sensors are leveraged to locate the position of the UAS, so that the UAS could navigate itself to the destination. The image matching technology enable the UAS to leverage the cameras to get the destinations. The laser sensors are also a powerful tool to get location and navigation for the UAS. Novel approaches are needed to interfere and spoof these sensors or invalid these functions to force the UAS to get back.
    \item The UAS controlled by radio communicate with each end device with protocols. There is a need to recognize the protocol used by the autopilot and communication devices. Because this approach could enable the officials to determine the parameters of the communication and attack the command link used by the pilots on the ground. At the same time, the recognition approach could be executed in the SDR so that the surveillance UAS could leverage the SDR to decrypt the communication packet and modify the flight configuration to return.
\end{itemize}

\section{Future Trends}
\subsection{Technical advancements}
As discussed above, simple detection approaches cannot get a reliable rate of detecting malicious UAS successfully. On one hand, each simple approach has its disadvantages so the simple detection sensor could not meet all the detection requirements in a variable environment; on the other hand, the UAS designed with different materials and configurations also pose a big challenge for simple detection sensors to capture. The future UAS detection approaches would be more mature, practical and efficient. The detection schemes need to be combined from multiple sensors and fused with ground data and aerial data collaboratively. The diversity and the amount of data in types of detection and acquisition space could be a trend to improve. Similar to detection schemes, simple negation approaches also could not satisfy negation requirement, especially the countermeasures could not damage the property of the pilots. Future negation schemes should focus on navigating the intruding UAS to fly away the sensitive areas and no harm to the property of pilots. Of course, the different countermeasures are promising to achieve better performance when they are implemented collaboratively. That means how to construct a unified and systematic framework for the UAS safety defense also is a challenge in the following stages. What's more, a UAS safety defense system includes detection and negation, so how to balance the two parts in a collaborative and unified framework is also a research focus. 
\subsection{Industrial Standards}
 Most security and safety problems are caused by the people's mistaking operation. The detailed operation and management policy could be helpful to people to avoid the mistakes and reduce the burden of the defenders. These policies not only serve as guidance to the pilots, but also standards to the industries. A guidance allows the pilots to make awareness of flight of UAS in safety and security and avoid the mistaking operations when the UAS is on the flight. The industrial standards make it possible to stop the UAS when it is out of control. 
 \begin{itemize}
     \item The industrial standards need the market entrance standards which limit the UAS on the market to be controllable and identified in a physical level. Once the UAS are instructing in restricted areas, the defenders could be able to access the system in physical level to drive away or stop the UAS remotely.
     \item The basic training for pilots needs to include the safety operations and security knowledge. The specification of UAS operation training and certification could help pilots avoid basic mistakes. 
 \end{itemize} 
\subsection{Unified and Secured Coordination Strategies}
The discussion above shows that it is hard to protect the public from unsafe and unauthorized drone operations by using one single approach. Therefore, we propose a unified framework of collaborative UAS safety management, as shown in Fig. \ref{figUnifiedFramework}. The collaborated entities for UAS safety management are:
\begin{enumerate}
    \item \textbf{Local UAS coordinator:} The local UAS authorities \cite{FAA_UAS} are responsible for making use of all interfaces provided by UAS manufactures to secure the operation of UAS and handing over UAS within coordinator when necessary. Obviously this is a distributed management paradigm \cite{DHS}.
    \item \textbf{Existing airspace authorities:} Existing airspace management authorities are not required to deal with UAS directly. It is desired for airspace management authorities to interact with the regulation database and release information to the shared data cube \cite{Sightings,FAA}, which is supposed to shared with local UAS coordinators friendly. The key information can be visualized to figure out the status of UAS management quickly and effectively. Apart from the data sharing, the authorities are supposed to maintenance the security and the integrity of the releasing data in case modified by attackers. Right before the publication of this paper, we do notice that FAA has released a mobile App to graphically display where drone operation is allowed as well as specific rules to follow \cite{B4UFLY}.
    \item \textbf{UAS manufacturers:} Manufacturers are responsible for specifying the minimal environmental requirement for proper manipulation \cite{SecuritySensitive} of UAS. Meanwhile, they are required to provide privileged control interfaces for local UAS coordinators \cite{GAO} to interrupt and re-accommodate UAS when necessary. Detect and avoid (DAA) technologies will play an important role in overcoming barriers to UAS integration \cite{SA1,SA2,SA3,SA4,SA5,SA6}.
    \item \textbf{Local UAS coordinators: } This entity interacts as an agent between UAS users and airspace authorities, they are responsible for providing proper guidelines for UAS users and make use of privileged control interfaces to assure that UAS operations comply with issued regulations.
    \item \textbf{Trusted information provider:} The information providers are responsible for: a) providing the whole framework with safety and security related information. b) reviewing the report submitted from residents \cite{COMMAG18}.
\end{enumerate}
Within our proposed framework, UAS operations are within the supervision of local UAS coordinators and UAS manufactures. The residents could achieve a more secure and safe life under a harmony management of the UAS traffic environment. 

\section{Concluding Remarks}
In addition to recreational use, unmanned aircraft systems (UAS), also known as unmanned aerial vehicles (UAV) or drones, are used across our world to support firefighting and search and rescue operations, to monitor and assess critical infrastructure, to provide disaster relief by transporting emergency medical supplies to remote locations, and to aid efforts to secure our borders. However, UAS can also be used for malicious schemes by terrorists, criminal organizations, and lone actors with specific objectives. To promote safe, secure and privacy-respecting UAS operations, there is an urgent need for innovative technologies for detecting and mitigating UAS. Over the past 5 years, significant research efforts have been made to counter UAS: detection technologies are based on acoustic, vision, passive radio frequency, or data fusion; and mitigation technologies include physical capture or jamming. In this paper, we provided a comprehensive survey of existing literature in the area of UAS detection and mitigation, identified the challenges in countering unauthorized or unsafe UAS, and evaluated the trends of detection and mitigation for protecting against UAS-based threats. We envision that an integrated system capable of detecting and mitigating UAS will be essential to the safe integration of UAS into the airspace system. 
    
\section*{Acknowledgment}
This research was partially supported through Embry-Riddle Aeronautical University's Faculty Innovative Research in Science and Technology (FIRST) Program and the National Science Foundation under grant No. 1956193.

\bibliographystyle{IEEEtran}
\bibliography{main}

\begin{IEEEbiography}[{\includegraphics[width=1in,height=1.25in,clip,keepaspectratio]{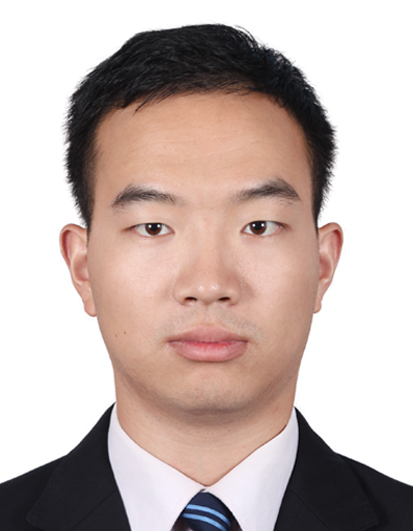}}]{Jian Wang}
(wangj14@my.erau.edu) is a Ph.D. student in the Department of Electrical Engineering and Computer Science, Embry-Riddle Aeronautical University (ERAU), Daytona Beach, Florida, and a graduate research assistant in the Security and Optimization for Networked Globe Laboratory (SONG Lab, www.SONGLab.us). He received his M.S. from South China Agricultural University (SCAU) in 2017 and B.S. from Nanyang Normal University in 2014. His major research interests include wireless networks, unmanned aerial systems, and machine learning. He was a recipient of the Best Paper Award from the 12th IEEE International Conference on Cyber, Physical and Social Computing (CPSCom-2019).

\end{IEEEbiography}

\begin{IEEEbiography}[{\includegraphics[width=1in,height=1.25in,clip,keepaspectratio]{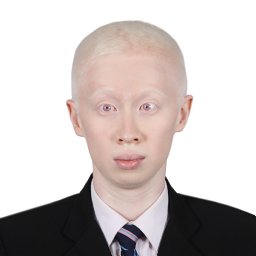}}]{YongXin Liu}
(LIU11@my.erau.edu) received his B.S. and M.S. from SCAU in 2011 and 2014, respectively, and he received Ph.D. from the School of Civil Engineering and Transportation, South China University of Technology. His major research interests include data mining, wireless networks, the Internet of Things, and unmanned aerial vehicles. He was a recipient of the Best Paper Award from the 12th IEEE International Conference on Cyber, Physical and Social Computing (CPSCom-2019).
\end{IEEEbiography}

\begin{IEEEbiography}[{\includegraphics[width=1in,height=1.25in,clip,keepaspectratio]{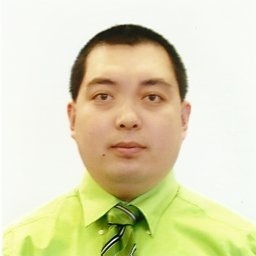}}]{Houbing Song} (M'12-SM'14) received the Ph.D. degree in electrical engineering from the University of Virginia, Charlottesville, VA, in August 2012, and the M.S. degree in civil engineering from the University of Texas, El Paso, TX, in December 2006.

In August 2017, he joined the Department of Electrical Engineering \& Computer Science, Embry-Riddle Aeronautical University, Daytona Beach, FL, where he is currently an Assistant Professor and the Director of the Security and Optimization for Networked Globe Laboratory (SONG Lab, www.SONGLab.us). He served on the faculty of West Virginia University from August 2012 to August 2017. In 2007 he was an Engineering Research Associate with the Texas A\&M Transportation Institute. In 2019 he served as an AI and Counter Cyber for Autonomous Unmanned Collective Control Subject Matter Expert selected by the United States Special Operations Command (USSOCOM). He has served as an Associate Technical Editor for IEEE Communications Magazine (2017-present), an Associate Editor for IEEE Internet of Things Journal (2020-present) and a Guest Editor for IEEE Journal on Selected Areas in Communications (J-SAC), IEEE Internet of Things Journal, IEEE Transactions on Industrial Informatics, IEEE Sensors Journal, IEEE Transactions on Intelligent Transportation Systems, and IEEE Network. He is the editor of six books, including Big Data Analytics for Cyber-Physical Systems: Machine Learning for the Internet of Things, Elsevier, 2019,  Smart Cities: Foundations, Principles and Applications, Hoboken, NJ: Wiley, 2017, Security and Privacy in Cyber-Physical Systems: Foundations, Principles and Applications, Chichester, UK: Wiley-IEEE Press, 2017, Cyber-Physical Systems: Foundations, Principles and Applications, Boston, MA: Academic Press, 2016, and Industrial Internet of Things: Cybermanufacturing Systems, Cham, Switzerland: Springer, 2016.  He is the author of more than 100 articles. His research interests include cyber-physical systems, cybersecurity and privacy, internet of things, edge computing, AI/machine learning, big data analytics, unmanned aircraft systems, connected vehicle, smart and connected health, and wireless communications and networking. His research has been featured by popular news media outlets, including IEEE GlobalSpec's Engineering360, USA Today, U.S. News \& World Report, Fox News, Association for Unmanned Vehicle Systems International (AUVSI), Forbes, WFTV, and New Atlas.

Dr. Song is a senior member of ACM. Dr. Song was a recipient of the Best Paper Award from the 12th IEEE International Conference on Cyber, Physical and Social Computing (CPSCom-2019), the Best Paper Award from the 2nd IEEE International Conference on Industrial Internet (ICII 2019), the Best Paper Award from the 19th Integrated Communication, Navigation and Surveillance technologies (ICNS 2019) Conference, and the prestigious Air Force Research Laboratory's Information Directorate (AFRL/RI) Visiting Faculty Research Fellowship in 2018. 
\end{IEEEbiography}






\end{document}